\documentclass[11pt,a4paper,reqno]{amsart}
\usepackage[utf8]{inputenc}
\usepackage[T1]{fontenc}
\usepackage{mathptmx}
\usepackage{extarrows}
\usepackage{amsmath}
\usepackage{amssymb}
\usepackage{amsthm}
\usepackage{amsxtra}
\usepackage[cal=boondoxo, calscaled=1.1]{mathalfa}
\usepackage{upgreek}
\usepackage{bbm}
\usepackage{mathtools}
\usepackage{stmaryrd}

\usepackage{extarrows}
\usepackage{enumitem}
\usepackage{tikz-cd}
\usepackage{url}
\usepackage{booktabs}
\usepackage{graphicx}
\usepackage{shuffle}
\usepackage{resizegather}
\usepackage{arydshln}
\usepackage{floatrow}
\usepackage{soul}
\usepackage{relsize,etoolbox}
 \usepackage[all,cmtip,2cell]{xy}
\UseAllTwocells
\xyoption{2cell}

 \numberwithin{equation}{section}

 \usepackage[pdftex,
                paper=a4paper,
                portrait=true,
                textwidth=165mm,
                textheight=238mm,
                tmargin=2.5cm,
                marginratio=1:1]{geometry}

\theoremstyle{definition}

\theoremstyle{remark}

 \newtheorem*{note}{Notation}

\def\ud{\mathrm{d}}
\def\td{\mathrm{t.d.}}
\def\cs{\mathrm{cs}}
\def\cd{\mathrm{cd}}
\def\uW{\mathrm{W}}
\def\id{\mathrm{id}}

\newcommand{\ui}{\mathrm{i}\mkern1mu}

\newcommand{\RR}{\ensuremath{\mathbbmss{R}}}

\newcommand{\ucal}{\mathcal{u}}
\newcommand{\vcal}{\mathcal{v}}
\newcommand{\hcal}{\mathcal{h}}

\newcommand{\Acal}{\mathcal{A}}

\newcommand{\Ecal}{\mathcal{E}}

\newcommand{\Mcal}{\mathcal{M}}

\newcommand{\Wcal}{\mathcal{W}}

\newcommand{\OWcal}{\mathcal{O}\mathcal{W}}

\newcommand{\gfrak}{\mathfrak{g}}
\newcommand{\Lfrak}{\mathfrak{L}}
\newcommand{\Cfrak}{\mathfrak{C}}
\newcommand{\Scal}{\mathfrak{Scal}}
\newcommand{\Kin}{\mathfrak{Kin}}

\newcommand{\kbf}{\mathbf{k}}
\newcommand{\ubf}{\mathbf{u}}
\newcommand{\xbf}{\mathbf{x}}

\newcommand{\End}{\operatorname{End}}
\newcommand{\grad}{\operatorname{\mathbf{grad}}}
\newcommand{\rot}{\operatorname{\mathbf{rot}}}
\renewcommand{\div}{\operatorname{\mathbf{div}}}
\newcommand{\coker}{\operatorname{coker}}

\newcommand{\ue}{\operatorname{e}}

\newcommand{\lbbar}{\{\kern-0.76ex\{}
\newcommand{\rbbar}{\}\kern-0.76ex\}}

\newcommand*{\sbullet}{\raisebox{0.1ex}{\scalebox{0.6}{$\bullet$}}}

\let\originalleft\left
\let\originalright\right
\renewcommand{\left}{\mathopen{}\mathclose\bgroup\originalleft}
\renewcommand{\right}{\aftergroup\egroup\originalright}

\DeclareMathAlphabet{\mathbfit}{OML}{cmm}{b}{it}

\newcommand*\oline[1]{%
  \vbox{%
    \hrule height 0.5pt
    \kern0.25ex
    \hbox{%
      \kern-0.1em
      \ifmmode#1\else\ensuremath{#1}\fi
      \kern-0.1em
    }
  }
}

\title[Homotopy double copy for the non-abelian and tensor Navier-Stokes equations]{Homotopy double copy and the Kawai-Lewellen-Tye relations for the non-abelian and tensor Navier-Stokes equations}

\author{Valentina Guar\'{i}n Escudero} 
\address{Escuela de Matem\'{a}ticas\\ Universidad Nacional de Colombia Sede Medell\'{i}n \\ Carrera 65 $\#$ 59A--110 \\ Medell\'{i}n \\ Colombia}
\email{mvguarine@unal.edu.co}

\author{Cristhiam Lopez-Arcos} 
\address{Escuela de Matem\'{a}ticas\\ Universidad Nacional de Colombia Sede Medell\'{i}n \\ Carrera 65 $\#$ 59A--110 \\ Medell\'{i}n \\ Colombia}
\email{cmlopeza@unal.edu.co}

\author{Alexander Quintero V\'{e}lez} 
\address{Escuela de Matem\'{a}ticas\\ Universidad Nacional de Colombia Sede Medell\'{i}n \\ Carrera 65 $\#$ 59A--110 \\ Medell\'{i}n \\ Colombia}
\email{aquinte2@unal.edu.co}

\begin{document}

\dedicatory{To the memory of Jorge Alberto Naranjo Mesa}

\begin{abstract}
Recently, a non-abelian generalisation of the Navier-Stokes equation that exhibits a manifest duality between colour and kinematics has been proposed by Cheung and Mangan. In this paper, we offer a new perspective on the double copy formulation of this equation, based on the homotopy algebraic picture suggested by Borsten, Kim, Jur\v{c}o, Macrelli, Saemann, and Wolf. In the process, we describe precisely how the double copy can be realised at the level of perturbiner expansions. Specifically, we will show that the colour-dressed Berends-Giele currents for the non-abelian version of the Navier-Stokes equation can be used to construct the Berends-Giele currents for the double copied equation by replacing the colour factors with a second copy of kinematic numerators. We will also show a Kawai-Lewellen-Tye relation stating that the full tree-level scattering amplitudes in the latter can be written as a product of tree-level colour ordered partial amplitudes in the former. 
\end{abstract}

 \maketitle

\section{Introduction}
In 1985, Kawai, Lewellen and Tye~\cite{Kawai:1985xq} discovered an astonishing relation between tree-level closed and open string amplitudes, which after taking the field theory limit implies that the tree-level gravity amplitudes can be represented by the ``square’’ of the tree-level colour-ordered Yang-Mills amplitudes. It was only realised much later by Bern, Carrasco, and Johansson~\cite{Bern:2008qj} that if the Yang-Mills amplitudes can be arranged so that their numerator factors satisfy a specific color-kinematics duality, it is possible to directly generate gravity amplitudes from these by squaring the numerators. This method for obtaining gravity amplitudes is known as the double copy construction (see Ref.~\cite{Bern:2019prr} for a review). 

Since its original formulation, it has become clear that the double copy construction is not restricted to just gravity and Yang-Mills theory. Identical squaring relations hold for a much larger ``web’’ of theories. Notable examples include the non-linear sigma model, the special Galileon, and Born-Infeld theories~\cite{Chen:2013fya,Cachazo:2014xea,Cheung:2014dqa,Du:2016tbc,Cheung:2016prv,Cheung:2017ems}. Even more, it turns out that the double copy can also be used to engineer scattering amplitudes in nonrelativistic theories, emphasising the widespread applicability of these ideas. Let us comment briefly on this, as it will be our main focus.

In a recent article \cite{Cheung:2020djz}, Cheung and Mangan initiated the study of scattering amplitudes for a natural non-abelian generalisation of the Navier-Stokes equation, which they dubbed the non-abelian Navier-Stokes equation. Essentially following the steps of the analysis given in Ref.~\cite{Wyld:1961gqg}, these authors derived the Feynman rules for the perturbative scheme to this equation and used them to calculate the tree-level amplitudes for the scattering of three and four fluid quanta. They also showed that the tree-level amplitude for the scattering of an arbitrary number of fluid quanta is ``on-shell constructible'' by carefully examining its infrared properties. But perhaps more significant was their observation that for any triplet of off-shell Feynman diagrams describing the $s$, $t$ and $u$ channel fluid quanta exchanges, the kinematic Jacobi identities required for colour-kinematics duality to hold are automatically satisfied. This enabled them to apply the double copy procedure at the classical level to ``square'' the non-abelian Navier-Stokes equation, thus obtaining the tensor Navier-Stokes equation that governs the dynamics of a bi-fluid velocity distribution. 

In the present paper, we will explore the duality between colour and kinematics and the double copy of the non-abelian Navier-Stokes equation in more depth. Our approach relies on the remarkable description of the double copy formalism in terms of factorisations of strict $L_{\infty}$-algebras introduced by Borsten~\emph{et.~al.} in Ref.~\cite{Borsten:2021hua}.  However, in contrast to the point of view taken in that reference, here we will be working with perturbiner expansions, which are multiparticle solutions to the non-abelian and tensor Navier-Stokes equations that arise in the construction of the minimal model of the strict $L_{\infty}$-algebras underlying these equations \cite{Lopez-Arcos:2019hvg,Gomez:2020vat}. This provides us with a unified picture for treating the various realisations of the double copy procedure that we will consider. 

To make things concrete, let us sketch the main qualitative results of this paper. We will first show that the strict $L_{\infty}$-algebra $\Lfrak$ that controls both the colour-stripped and the colour-dressed perturbiner expansions for the non-abelian Navier-Stokes equation factorises into a colour factor, a kinematic factor, and a scalar theory factor. To be precise,
\begin{equation*}
\Lfrak = \gfrak \otimes (\Kin \otimes_{\tau} \Scal),
\end{equation*}
where $\gfrak$ is the colour factor corresponding to the Lie algebra of internal symmetries, $\Kin$ is the kinematic factor corresponding to the kinematic degrees of freedom, and $\Scal$ is the scalar theory factor corresponding to the strict $L_{\infty}$-algebra that encodes the trivalent interactions. Furthermore, just as the general construction of Ref.~\cite{Borsten:2021hua} dictates, the tensor product between the last two factors is twisted by a twist datum $\tau$. Next we will verify that this factorisation of $\Lfrak$ is suitable, which means that it is compatible with colour-kinematics duality. This will be done by showing that it is possible to reorganise the colour-dressed Berends-Giele currents produced by the non-abelian Navier-Stokes equation so as to extract kinematic numerators that satisfy the same generalised Jacobi identities as their colour counterparts. As a matter of fact, we will identify the kinematic Lie algebra that gives rise to these numerators with the Lie algebra of ``Fourier coefficients'' of infinitesimal spatial diffeomorphisms, in complete agreement with the statement made in Ref.~\cite{Cheung:2020djz}. Once this is realised, we will demonstrate that the strict $L_{\infty}$-algebra $\Lfrak'$ controlling the perturbiner expansions for the tensor Navier-Stokes equation can be represented  by a ``homotopy double copy'', in the sense of Ref.~\cite{Borsten:2021hua}, of the strict $L_{\infty}$-algebra $\Lfrak$. What this means concretely is that, from the above factorisation of $\Lfrak$, we obtain $\Lfrak'$ by replacing the colour factor $\gfrak$ with another copy of the kinematic factor $\Kin$ and twist the tensor product by~$\tau$: 
\begin{equation*}
\Lfrak' = \Kin  \otimes_{\tau} (\Kin \otimes_{\tau} \Scal).
\end{equation*}
Within this context, we next address the question of how this factorisation can be practically implemented both at the level of perturbiner expansions and at the level of scattering amplitudes. For the former, we will show that the double copy Berends-Giele currents can be obtained from the colour-dressed Berends-Giele currents by simply substituting the colour factors for another copy of kinematic numerators. For the latter, we will show a Kawai-Lewellen-Tye relation that provides a representation of the tree-level scattering amplitudes of bi-fluid quanta described by the tensor Navier-Stokes equation as the ``square'' of tree-level colour ordered partial amplitudes of fluid quanta described by the non-abelian Navier-Stokes equation. Of course, as with the standard Kawai-Lewellen-Tye factorisation of gravity amplitudes into products of Yang-Mills amplitudes \cite{Cachazo:2013iea,Mafra:2016ltu,Mizera:2016jhj}, the role of the momentum kernel will be played by the inverse of a matrix which is constructed out of Berends-Giele double currents derived from a strict $L_{\infty}$-algebra $\Lfrak''$ that controls the perturbiner expansions of a bi-adjoint analogue of the Navier-Stokes equation. This strict $L_{\infty}$-algebra, in turn, will be shown to be represented by a ``homotopy zeroth copy'' of the the strict $L_{\infty}$-algebra $\Lfrak$, again in the spirit of Ref.~\cite{Borsten:2021hua}. In other words, $\Lfrak''$ is obtained by replacing the kinematic factor $\Kin$ with a different colour factor $\bar{\gfrak}$ in the prescribed factorisation of $\Lfrak$:
\begin{equation*}
\Lfrak'' = \gfrak \otimes (\bar{\gfrak} \otimes \Scal).
\end{equation*}
To summarise, we will find that the homotopy algebraic perspective on the double copy advocated in Ref.~\cite{Borsten:2021hua} gives us a complete and elegant formulation of this operation at all levels for the non-abelian Navier-Stokes equation. 
 
It is significant to point out that the scalar factor $\Scal$ is common to all three strict $L_{\infty}$-algebras: $\Lfrak$, $\Lfrak'$ and $\Lfrak''$. Since, as we have already stated, the strict $L_{\infty}$-algebra $\Scal$ describes a  scalar theory with only trivalent interactions that underlies the non-abelian, tensor, and bi-adjoint Navier-Stokes equations, which are manifestly nonrelativistic, this means that none of these three equations can be obtained from the extremisation of an action functional.\footnote{The overall reason for this is that the nonrelativistic scalar field encased in $\Scal$ is necessarily complex, and it is therefore not possible to write a suitable action functional with a cubic interaction. It is also worth noting here that this fact is not too surprising when we recall that the classical Navier-Stokes equation, of which the equations in question are generalisations, does not admit a least action principle.} What this translates to from the algebraic standpoint is that none of the three strict $L_{\infty}$-algebras $\Lfrak$, $\Lfrak'$ and $\Lfrak''$ can be equipped with a cyclic inner product giving rise to a homotopy Maurer-Cartan action; see Ref.~\cite{Jurco:2018sby} for an explanation. 

One motivation for this investigation is that the non-abelian Navier-Stokes equation is rich enough to elucidate the general structure of the double copy prescription, avoiding many of the technical complications that arise in the generic situation, where a specific strictification compatible with colour-kinematics duality of the $L_{\infty}$-algebra describing the dynamics needs to be singled out before one can actually apply the recipe developed in Ref.~\cite{Borsten:2021hua}. Another motivation is to show how perturbiner methods will help give perspective and a deeper understanding of the results in \emph{loc.~cit.} This is fully in line with the arguments and findings presented in Ref.~\cite{Ahmadiniaz:2021ayd}, where it was confirmed that our proposal for determining the double copied perturbiner coefficients works out for several other theories. 

The organisation of this paper is as follows. In Section~\ref{sec:2}, we discuss the non-abelian Navier-Stokes equation and outline some of its basic properties. After these preparations, we explain the construction of the strict $L_{\infty}$-algebra $\Lfrak$ relevant to such equation in Section~\ref{sec:3}. In Sections~\ref{sec:4} and~\ref{sec:5}, we determine multiparticle solutions to the non-abelian Navier-Stokes equation in the form of colour-stripped and colour-dressed perturbiner expansions. In Section~\ref{sec:6}, we tackle the task of factorising the strict $L_{\infty}$-algebra $\Lfrak$, so that the construction described above can be carried out. Section~\ref{sec:7} is devoted to the verification of the fact that this factorisation is compatible with colour-kinematics duality. In Section~\ref{sec:8} we introduce the strict $L_{\infty}$-algebra $\Lfrak'$ associated to the tensor Navier-Stokes equation and show that it can be obtained as the homotopy double copy of $\Lfrak$. In Section~\ref{sec:9} we work out the multiparticle solution to the tensor Navier-Stokes equation in the shape of a perturbiner expansion. Section~\ref{sec:10} is concerned with the precise relation between the double copy Berends-Giele currents and the colour-dressed Berends-Giele currents. In Section~\ref{sec:11} we introduce the strict $L_{\infty}$-algebra $\Lfrak''$ that describes the bi-adjoint Navier-Stokes equation and show that it can be characterised as the homotopy zeroth copy of $\Lfrak$. Section~\ref{sec:12} is dedicated to the determination of multiparticle solutions to the bi-adjoint Navier-Stokes equation by way of colour-stripped and colour-dressed perturbiner expansions. Following this, scattering amplitudes for the non-abelian and tensor Navier-Stokes equations are discussed on Section~\ref{sec:13}. We then proceed in Section~\ref{sec:14} to show the Kawai-Lewellen-Tye relations. Finally, in Section~\ref{sec:15} we offer our conclusions and speculations.


\begin{note}
The physical space will be represented isometrically by the $3$-dimensional Euclidean space $\RR^3$. 
To specify its points we use a rectangular coordinate system $(x^{1},x^{2},x^{3})$. We refer to the coordinate triple $(x^{1},x^{2},x^{3})$ as the position and denote it by $\xbf$.  Latin indices $i$, $j$, etc. run over the coordinate labels $1,2,3$. The parameter $t$ is identified with the time, and we may suppose its range to be $-\infty < t < \infty$. The coordinate basis corresponding to $(x^{i})$ is denoted $e_{i}$, and the dual basis is denoted $e^{i}$. The terminology  ``vector'' is reserved for $e_{i}$ and the terminology ``$1$-form'' is used for $e^{i}$. We also adopt the shorthand notation $\partial_{i}$ for the partial derivative $\frac{\partial}{\partial x^{i}}$.

We use $\Omega_{\td}^{r}(\RR^3)$ to denote the space of time-dependent $r$-forms on $\RR^3$. We also write $\ud \colon \Omega_{\td}^{r}(\RR^3) \to \Omega_{\td}^{r+1}(\RR^3)$ for the spatial exterior differential, $\ast \colon \Omega_{\td}^{r}(\RR^3) \to \Omega_{\td}^{3-r}(\RR^3)$ for the spatial star operator with respect to the usual metric on $\RR^3$, $\delta  = \ast \ud  \ast \colon \Omega_{\td}^{r}(\RR^3) \to \Omega_{\td}^{r-1}(\RR^3)$ for the corresponding spatial exterior codifferential, and $\frac{\partial}{\partial t} \colon \Omega_{\td}^{r}(\RR^3) \to \Omega_{\td}^{r}(\RR^3)$ for the partial derivative with respect to time.

If $\ubf$ is a time-dependent vector field on $\RR^3$, we use the symbol $\ubf^{\flat}$ to denote the associated time-dependent $1$-form on $\RR^3$. In terms of the coordinate system $(x^{i})$, the components of $\ubf^{\flat}$ are obtained by lowering the indices of the components of $\ubf$. Of course, we can also pass from a time-dependent $1$-form $u$ to a time-dependent vector field $u^{\sharp}$ by raising the indices of the components of $u$. This enables us to define a Lie bracket on the space $\Omega_{\td}^{1}(\RR^3)$ by setting
\begin{equation*}
[u,v] = [u^{\sharp},v^{\sharp}]^{\flat},
\end{equation*}
where the bracket on the right-hand side is just the Lie bracket of two time-dependent vector fields on $\RR^3$. If we write this out explicitly in components, we have
\begin{equation*}
[u,v]_{i} = u^{j} \partial_{j} v_{i} - v^{j} \partial_{j} u_{i}.
\end{equation*}
The Einstein summation convention where repeated indices implies addition has been employed in this formula and, in the sequel we shall adhere to this notation. 

We also frequently make use of a notation based on the combinatorics of words. By a word we mean a finite string $P=p_1p_2\cdots p_k$  of positive integers $p_1,p_2,\dots,p_k \geq 1$. The word consisting of no symbols, called the empty word, is represented by $\varnothing$. Given a word $P= p_1p_2\cdots p_k$, we denote by $\bar{P}=p_k p_{k-1} \cdots p_1$ its transpose and by  $\lvert P \rvert$ its length $k$. The scalar product of two words $P$ and $Q$ is defined to be
\begin{equation*}
(P,Q) = \delta_{P,Q},
\end{equation*}
where $\delta_{P,Q}$ is the `Kronecker delta', that is, $\delta_{P,Q} = 1$ if $P = Q$ and $\delta_{P,Q} = 0$ if $P \neq Q$. Clearly, this definition may be extended by linearity to arbitrary finite linear combinations of words. We too shall need the shuffle product on words, which will be denoted by the symbol $\shuffle$. It is defined inductively by setting 
\begin{equation*}
\varnothing \shuffle P = P \shuffle \varnothing = P, \quad pP \shuffle qQ  = p(P \shuffle q Q) + q(pP \shuffle Q), 
\end{equation*}
for any words $P$ and $Q$ and for any positive integers $p$ and $q$. 

We shall implicitly work with the free Lie algebra $\mathcal{L}_{n-1}$ generated by the set $\{1,2,\dots, n-1\}$. This is defined as the span of the letters in $\{1,2,\dots, n-1\}$ and all brackets of letters in $\{1,2,\dots, n-1\}$. By a Lie polynomial of length $n-1$ we mean an element of $\mathcal{L}_{n-1}$. The left to right bracketing of a word $P = p_1 p_2 \cdots p_k$ with letters in $\{1,2,\dots, n-1\}$ is denoted $\ell[P] = [[\dots[[p_1,p_2],p_3],\dots],p_k]$. Any Lie polynomial $\Gamma$ of length $n-1$ may be expanded as
\begin{equation*}
\Gamma = \sum_P (1P, \Gamma) \ell[1P],
\end{equation*}
where the sum ranges over all permutations $P$ of $23 \cdots (n-1)$. 

Finally we shall need the notion of an object labelled by words satisfying the generalised Jacobi identities. Specifically, let $U_P$ be an arbitrary object which is labelled by words $P$ with letters in $\{1,2,\dots,n-1\}$. Then $U_P$ is said to satisfy the generalised Jacobi identity of order $k$ if
\begin{equation*}
U_{Q \ell[R]} + U_{R \ell[Q]} = 0
\end{equation*}
for every pair of non-empty words $Q$ and $R$ such that $\lvert Q \rvert + \lvert R \rvert = k$. In this case, we shall write $U_{\ell[P]}$ instead of $U_P$. In particular, this implies that $U_{[\ell[Q],\ell[R]]}= U_{\ell[Q\ell[R]]}$.
\end{note}


\section{Non-abelian Navier-Stokes equations}\label{sec:2}
We shall start in this section by discussing the non-abelian analogue of the Navier-Stokes equations. Such analogue has been introduced recently by Cheung and Mangan in Ref.~\cite{Cheung:2020djz}. However, our presentation differs from theirs in that we follow an intrinsic, and basically coordinate free approach. We begin by reviewing the main ingredients associated with the classical Navier-Stokes equations. 

The mathematical description of the state of a homogeneous and incompressible Newtonian fluid with unit density is effected by means of a vector field $\ubf = \ubf(\xbf,t)$ which gives the distribution of the fluid  velocity and a function $p=p(\xbf,t)$ which gives the pressure. Its dynamics is governed by the Navier-Stokes equations expressed as
\begin{subequations}
\begin{align}   
&\div \ubf = 0, \label{eq:2.1a}\\
&\frac{\partial \ubf}{\partial t} - \nu \Delta \ubf + (\ubf \cdot \grad)\ubf + \grad p = 0, \label{eq:2.1b}
\end{align}
\end{subequations}
where $\nu$ is the kinematic viscosity. Using the formulas well known in vector analysis, $\Delta\ubf = \grad(\div \ubf) - \rot (\rot \ubf)$ and $(\ubf \cdot \grad)\ubf  =\frac{1}{2} \grad  (\ubf \cdot \ubf) - \ubf \times \rot \ubf$, and the incompressibility condition \eqref{eq:2.1a}, we may put Eq.~\eqref{eq:2.1b} in the form
\begin{equation}\label{eq:2.2}
\frac{\partial \ubf}{\partial t} + \nu \rot (\rot \ubf) + \tfrac{1}{2} \grad  (\ubf \cdot \ubf) - \ubf \times \rot \ubf + \grad p = \mathbf{0}.
\end{equation}
When the velocity distribution is known, the pressure distribution in the fluid can be found by solving the Poisson-type equation
\begin{equation}\label{eq:2.3}
\Delta p = - \tfrac{1}{2}\Delta(\ubf \cdot \ubf) + \div (\ubf \times \rot \ubf),
\end{equation}
which is obtained by taking the divergence of Eq.~\eqref{eq:2.2}.

In order to derive a non-abelian version of the Navier-Stokes equations it is convenient to first translate Eqs.~\eqref{eq:2.1a} and \eqref{eq:2.2} into the language of time-dependent differential forms. To this end, we use the identities
\begin{align*}
\begin{split}
\left(\div \ubf \right)^{\flat} &= \delta \ubf^{\flat}, \\
\left(\rot (\rot \ubf)\right)^{\flat} &= \delta\ud \ubf^{\flat}, \\
\left( \grad  (\ubf \cdot \ubf)\right)^{\flat} &= \ud\!\ast \!(\ubf^{\flat} \wedge \ast \ubf^{\flat}), \\
\left(\ubf \times \rot \ubf\right)^{\flat} &= \ast (\ubf^{\flat} \wedge \ast \ud \ubf^{\flat}),\\
\left( \grad p \right)^{\flat} &= \ud p,
\end{split}
\end{align*}
where the notation is the one described in the introduction. Substituting these formulae in Eqs.~\eqref{eq:2.1a} and \eqref{eq:2.2}, the Navier-Stokes equations are then expressed as
\begin{subequations}
\begin{align}   
 &\delta \ubf^{\flat} = 0,\label{eq:2.4a}\\
&\frac{\partial \ubf^{\flat}}{\partial t} + \nu \delta  \ud \ubf^{\flat} +\tfrac{1}{2}\ud\!\ast\! (\ubf^{\flat} \wedge \ast \ubf^{\flat}) - \ast (\ubf^{\flat} \wedge \ast \ud \ubf^{\flat}) +  \ud p = 0.  \label{eq:2.4b}
\end{align}
\end{subequations}
When written in this fashion, the distribution of the fluid velocity is now represented by the time-dependent $1$-form $\ubf^{\flat}$, and the pressure by the time-dependent $0$-form $p$. 

Having this, we can formulate what we mean by a homogeneous and  incompressible non-abelian Newtonian fluid. We begin with a colour gauge group, which we take to be a compact Lie group $G$ with Lie algebra $\gfrak$. We are given a time-dependent $1$-form $u$ on $\RR^3$ with values in $\gfrak$, representing the distribution of the fluid velocity, and a time-dependent $0$-form $p$ on $\RR^3$ with values in $\gfrak$, representing the pressure. Taking inspiration from the universal form that the Navier-Stokes equations assume in the language of time-dependent differential forms, Eqs.~\eqref{eq:2.4a} and \eqref{eq:2.4b}, one may be tempted to stipulate that its non-abelian version is obtained by changing the wedge product by the Lie bracket on the space of time-dependent $r$-forms on $\RR^3$ taking values in $\gfrak$. However, this recipe does not reproduce the correct interaction term in the fluid velocity. To abstract such term properly, we must make a more educated guess. But first a little notation. 

Let $T_{a}$ be the generators of $\gfrak$, with structure constants $f_{ab}^{\phantom{ab}c}$ satisfying $[T_{a},T_{b}] = f_{ab}^{\phantom{ab}c} T_c$. Then an element $u$ of the space of time-dependent $1$-forms on $\RR^3$ with values in $\gfrak$ is specified by giving time-dependent $1$-forms $u^{a}$ on $\RR^3$. This makes it possible to define a binary operation on such space by means of
\begin{equation}\label{eq:2.5}
\lbbar u, v \rbbar = \lambda f_{bc}^{\phantom{bc}a} [u^{b}, v^{c}] T_{a}.
\end{equation}
Here $[,]$ stands for the Lie bracket on the space of time-dependent $1$-forms on $\RR^{3}$ and the coupling constant $\lambda$ has been inserted for later convenience. Written out in components, this equals
\begin{equation}\label{eq:2.6}
\lbbar u, v \rbbar^{a}_{i}  = \lambda f_{bc}^{\phantom{bc}a} (u^{bj} \partial_{j} v^{c}_{i} - v^{cj}\partial_{j}u^{b}_{i}).   
\end{equation}
That the operation so defined is independent of the choice of the generators $T_{a}$ is clear. 

Armed with this implement, we may now write down the non-abelian Navier-Stokes equations. They read
\begin{subequations}
\begin{align}   
 &\delta u = 0,\label{eq:2.7a}\\
&\frac{\partial u}{\partial t} + \nu \delta  \ud u + \tfrac{1}{2}\lbbar u, u \rbbar +  \ud p = 0.  \label{eq:2.7b}
\end{align}
\end{subequations}
To help understand these equations, we work out their component form. In the notation of the preceding paragraph, the components of $u$ and $p$ relative to the generators $T_{a}$ might be labeled as $u^{a}$ and $p^{a}$, respectively. Hence we may use Eq.~\eqref{eq:2.6} to rewrite Eqs.~\eqref{eq:2.7a} and~\eqref{eq:2.7b} in the forms
\begin{subequations}
\begin{align}   
 &\partial_{i} u^{ai} = 0,\label{eq:2.8a}\\
&\frac{\partial u^{a}_{i}}{\partial t} - \nu \Delta u^{a}_{i} + \lambda f_{bc}^{\phantom{bc}a} u^{bj} \partial_{j} u^{c}_{i} +  \partial_{i} p^{a} = 0.  \label{eq:2.8b}
\end{align}
\end{subequations}
These coincide precisely with the the sourceless non-abelian Navier-Stokes equations of Ref.~\cite{Cheung:2020djz}. The equations written in the form \eqref{eq:2.7a} and \eqref{eq:2.7b} have, however, a more intrinsic geometric meaning.

One other point we shall notice is this. Applying the spatial exterior codifferential $\delta$ to Eq.~\eqref{eq:2.7b} and using Eq.~\eqref{eq:2.7a} gives
\begin{equation}\label{eq:2.9}
\Delta p = - \tfrac{1}{2} \delta \lbbar u, u \rbbar. 
\end{equation}
In components, this implies that
\begin{equation}
\Delta p^{a} = - \lambda f_{bc}^{\phantom{bc}a} \partial_{i} (u^{bj} \partial_{j} u^{c i}) = - \lambda f_{bc}^{\phantom{bc}a} \partial_{i} u^{bj} \partial_{j} u^{c i},
\end{equation}
which is identically zero due to the antisymmetry of the structure constants $f_{bc}^{\phantom{bc}a}$ in their two lower indices. Thus Eq.~\eqref{eq:2.9} becomes simply
\begin{equation}
\Delta p = 0.
\end{equation}
From this we conclude that in a homogeneous and incompressible non-abelian Newtonian fluid, the velocity distribution $u$ and the pressure distribution $p$ are decoupled and may therefore be treated independently. This is in sharp contrast with the abelian situation, as evidenced by Eq.~\eqref{eq:2.3}. 

In view of the preceding remarks, we choose to ignore the pressure term in Eq.~\eqref{eq:2.7b} altogether, so that the non-abelian Navier-Stokes equations take the form
\begin{subequations}
\begin{align}   
 &\delta u = 0,\label{eq:2.12a}\\
&\frac{\partial u}{\partial t} + \nu \delta \ud u + \tfrac{1}{2}  \lbbar u, u \rbbar = 0, \label{eq:2.12b}
\end{align}
\end{subequations}
or, in terms of components,
\begin{subequations}
\begin{align}   
 &\partial_{i} u^{ai} = 0,\label{eq:2.13a}\\
&\frac{\partial u^{a}_{i}}{\partial t} - \nu \Delta u^{a}_{i} + \lambda f_{bc}^{\phantom{bc}a} u^{bj} \partial_{j} u^{c}_{i}  = 0.  \label{eq:2.13b}
\end{align}
\end{subequations}
We shall also find it convenient to treat Eq.~\eqref{eq:2.12a}, or its component counterpart Eq.~\eqref{eq:2.13a}, as a subsidiary solenoidal condition.


\section{The non-abelian Navier-Stokes strict $L_{\infty}$-algebra}\label{sec:3}
In this section we shall describe the strict $L_{\infty}$-algebra associated to the non-abelian Navier-Stokes equation \eqref{eq:2.12b}. This is the algebraic object that will enable us to construct multiparticle solutions via the perturbiner method advocated in Refs.~\cite{Lopez-Arcos:2019hvg} and~\cite{Gomez:2020vat}. 

Denote by $\Omega^{1}_{\td}(\RR^3,\gfrak)$ the space of time-dependent $1$-forms on $\RR^3$ with values in $\gfrak$. As a cochain complex, the non-abelian Navier-Stokes strict $L_{\infty}$-algebra $\Lfrak$ is
$$
\Omega^1_{\td}(\RR^3 ,\gfrak )[-1] \xrightarrow{\frac{\partial}{\partial t}+ \nu \delta \ud} \Omega^1_{\td}(\RR^3  ,\gfrak )[-2].
$$
Thus, $\Omega^1_{\td}(\RR^3,\gfrak )$ is concentrated in degrees $1$ and $2$, and the differential $l_1$ is the operator $\frac{\partial}{\partial t}+ \nu \delta \ud$ acting on $\Omega^1_{\td}(\RR^3,\gfrak )$. The bracket $l_2 \colon \Omega^1_{\td}(\RR^3,\gfrak )[-1]^{\otimes 2} \rightarrow \Omega^1_{\td}(\RR^3,\gfrak )[-2]$ is defined by
\begin{equation}\label{eq:3.1}
l_2 (u,v) = \lbbar u, v \rbbar.
\end{equation}
Recalling the definition \eqref{eq:2.5}, this is evidently skew-symmetric and since the graded Jacobi identity is trivially satisfied, it turns the graded vector space $\Lfrak = \Omega^1_{\td}(\RR^3 ,\gfrak )[-1] \oplus  \Omega^1_{\td}(\RR^3 ,\gfrak )[-2]$ into a strict $L_{\infty}$-algebra. From the definitions it follows immediately that, for all $u \in \Omega^1_{\td}(\RR^3 ,\gfrak )[-1]$,
$$
l_1 (u) + \tfrac{1}{2} l_2(u,u) = \frac{\partial u}{\partial t} + \nu \delta \ud u +\tfrac{1}{2}\lbbar u, v \rbbar,
$$
and hence the Maurer-Cartan equation for the strict $L_{\infty}$-algebra $\Lfrak$ reproduces exactly the non-abelian Navier-Stokes equation \eqref{eq:2.12b}.

Now, the definition of the strict $L_{\infty}$-algebra $\Lfrak$ requires a slight modification that allows to deal with perturbiner expansions. First we must establish some notation. Let $(a_{p})_{p\geq 1}$ be an infinite multiset of colour indices associated with the Lie algebra $\gfrak$ and let $(\kbf_p,\omega_p)_{p \geq 1}$ be an infinite set of pairs with $\kbf_p \in \RR^3$ and $\omega_p \in \RR$ and such that $\ui \omega_p + \kbf_p^2 = 0$ for each $p \geq 1$.  Let also $\Wcal_n$ be the set of words of length $n$. If $P = p_1 p_2 \cdots p_n$ is one such word, we put $T_{a_P}= T_{a_{p_1}}T_{a_{p_2}} \cdots T_{a_{p_n}}$, $\kbf_P = \kbf_{p_1} + \kbf_{p_2} + \cdots + \kbf_{p_n}$ and $\omega_P = \omega_{p_1} + \omega_{p_2} + \cdots + \omega_{p_n}$. We denote by $\Ecal^{0}_{\td}(\RR^3, \gfrak)$ the space of time-dependent formal series of the form
\begin{equation} \label{eq:3.2}
h(\xbf,t) = \sum_{n \geq  1} \sum_{P \in \Wcal_n} \hcal_{P} \ue^{\ui(\kbf_P \cdot \xbf + \omega_P t)} T_{a_P}, 
\end{equation}
and  by $\Ecal^{1}_{\td}(\RR^3, \gfrak)$ the space of time-dependent $1$-forms on $\RR^3$ with coefficients on $\Ecal^{0}_{\td}(\RR^3, \gfrak)$. Borrowing the terminology from Ref.~\cite{Lopez-Arcos:2019hvg}, elements of $\Ecal^{1}_{\td}(\RR^3, \gfrak)$ are called colour-stripped perturbiner ansatzs. 

We would next like to extend the operator $\frac{\partial}{\partial t}+ \nu \delta\ud$ to all of $\Ecal^{1}_{\td}(\RR^3, \gfrak)$. This can be achieved using the colour-dressed version of the perturbiner ansatzs. In terms of the rectangular coordinates of $\RR^3$, an element of $\Ecal^{1}_{\td}(\RR^3, \gfrak)$ may be represented as $u = u_{i}(\xbf,t) e^{i}$ where, in accord with Eq.~\eqref{eq:3.2},
the components $u_{i}(\xbf,t)$ are formal series of the form
\begin{equation}\label{eq:3.3}
u_{i}(\xbf,t) = \sum_{n \geq  1} \sum_{P \in \Wcal_n} \ucal_{i P} \ue^{\ui(\kbf_P \cdot \xbf + \omega_P t)} T_{a_P}.
\end{equation}
With this in mind, for each sequence of positive integers $p_1 < p_2 < \cdots < p_n$, we set
\begin{equation}\label{eq:3.4}
f^{a}_{p_1p_2\cdots p_n} = f_{a_{p_1} a_{p_2}}{}^{b}f_{b a_{p_3}}{}^{c} \cdots f_{d a_{p_{n-1}}}{}^{e} f_{e a_{p_{n}}}{}^{a},
\end{equation}
and define the coefficients $\ucal^{a}_{i p_1 p_2 \cdots p_n}$ by
\begin{equation}\label{eq:3.5}
\ucal^{a}_{i p_1 p_2 \cdots p_n} = \sum_{\sigma} f^{a}_{p_1p_{\sigma(2)}\cdots p_{\sigma(n)}} \ucal_{i p_1p_{\sigma(2)}\cdots p_{\sigma(n)}},
\end{equation}
the summation being taken over all permutations of the set $\{2,3,\dots, n\}$. With the help of the latter, Eq.~\eqref{eq:3.3} may be rewritten as $u_{i}(\xbf,t) = u_{i}^{a}(\xbf,t) T_{a}$, where the coefficients $u_{i}^{a}(\xbf,t)$ are formal series of the form
\begin{equation}
u_{i}^{a}(\xbf,t) = \sum_{n \geq 1} \sum_{P \in \OWcal_{n}} \ucal_{i P}^{a} \ue^{\ui (\kbf_{P} \cdot \xbf + \omega_P t)}.
\end{equation}
Here $\OWcal_{n}$ denotes the set of words $P=p_1p_2\cdots p_n$ of length $n$ with $p_1 < p_2 < \cdots < p_n$. This allows us to define the operator $\frac{\partial}{\partial t}+ \nu \delta\ud$ acting on the space $\Ecal^{1}_{\td}(\RR^3, \gfrak)$ by putting
\begin{equation}
\left( \frac{\partial}{\partial t}+ \nu \delta\ud\right) u_{i}^{a} (\xbf,t) =  \sum_{n \geq 1} \sum_{P \in \OWcal_{n}} (\ui \omega_P + \nu \kbf_P^2)\ucal_{i P}^{a} \ue^{\ui (\kbf_{P} \cdot \xbf + \omega_P t)}
\end{equation}
and extending by linearity. 

This digression out of the way, we can now write down the strict $L_{\infty}$-algebra $\Lfrak$ that conceals the perturbiner expansions for the non-abelian Navier-Stokes equation. The cochain complex underlying $\Lfrak$ is
\begin{equation}\label{eq:3.8}
\Ecal^1_{\td}(\RR^3 ,\gfrak )[-1] \xrightarrow{\frac{\partial}{\partial t}+ \nu \delta \ud} \Ecal^1_{\td}(\RR^3  ,\gfrak )[-2].
\end{equation}
Thus, as before, the differential $l_1$ is here the operator $\frac{\partial}{\partial t}+ \nu \delta\ud$ acting on $\Ecal^1_{\td}(\RR^3 ,\gfrak )$. As for the bracket $l_2 \colon \Ecal^1_{\td}(\RR^3,\gfrak )[-1]^{\otimes 2} \rightarrow \Ecal^1_{\td}(\RR^3,\gfrak )[-2]$, it is given by the same formula as that of Eq.~\eqref{eq:3.1}. That this is well-defined follows easily from the argument presented in \S2.2 of \cite{Mizera:2018jbh}.


\section{Colour-stripped multiparticle solution to the non-abelian Navier-Stokes equation}\label{sec:4}
Now we turn our attention to the multiparticle solution to the non-abelian Navier-Stokes equations. As we have already stated, this type of solution is obtained in the form of a perturbiner expansion \cite{Selivanov:1997aq, Rosly:1997ap,Mizera:2018jbh}. We take up first the case in which the perturbiner expansion is colour-stripped. The philosophy here is the same as that exposed in Ref.~\cite{Lopez-Arcos:2019hvg} in that the determination of such expansion can be reduced to the construction of a minimal model for the strict $L_{\infty}$-algebra $\Lfrak$.

To begin with, using the defining cochain complex \eqref{eq:3.8}, we see that the cohomology of $\Lfrak$ is concentrated in degrees $1$ and $2$. It is given by the solution space $H^{1}(\Lfrak)= \ker (l_1)$ of the linearaised equation $\frac{\partial u}{\partial t}+ \nu \delta \ud u = 0$ and the space $H^{2}(\Lfrak) = \coker (l_1)$ of linear on-shell colour-stripped perturbiner ansatzs. It follows that the cochain underlying the cohomology $H^{\sbullet}(\Lfrak)$ of $\Lfrak$ is
$$
\ker (l_1)[-1] \xrightarrow{\phantom{aa} 0 \phantom{aa}} \coker (l_1)[-2].
$$
In order to construct the minimal $L_{\infty}$-structure on $H^{\sbullet}(\Lfrak)$, we must define a projection $p \colon \Lfrak \to H^{\sbullet}(\Lfrak)$ and a contracting homotopy $h \colon \Lfrak \to \Lfrak$. To this end, we consider the Wyld propagator $G^{\uW}$ defined on the space of time-dependent $0$-forms on $\RR^{3}$. Its explicit expression, when acting on plane waves of the form $\ue^{\ui(\kbf \cdot \xbf + \omega t)}$, is
\begin{equation}\label{eq:4.1}
G^{\uW} = \frac{1}{ \ui \omega + \nu \kbf^2},
\end{equation}
as long as we assume that $\ui \omega + \nu \kbf^2 \neq 0$. We extend $G^{\uW}$ to all of $\Ecal^1_{\td}(\RR^3  ,\gfrak )$ so that we obtain a linear operator $G^{\uW} \colon \Ecal^1_{\td}(\RR^3  ,\gfrak ) \to \Ecal^1_{\td}(\RR^3  ,\gfrak )$ satisfying
\begin{equation}
l_1 \circ G^{\uW} = \left( \frac{\partial}{\partial t}+ \nu \delta \ud \right) \circ G^{\uW}  = \mathrm{id}_{\Ecal^1_{\td}(\RR^3  ,\gfrak )}.
\end{equation}
With the help of $G^{\uW}$, we may define the projection $p^{(1)} \colon \Ecal^1_{\td}(\RR^3  ,\gfrak ) \to \ker(l_1)$ by
\begin{equation}\label{eq:4.3}
p^{(1)} = \mathrm{id}_{\Ecal^1_{\td}(\RR^3  ,\gfrak )} - G^{\uW} \circ l_1.
\end{equation}
As for the other projection $p^{(2)} \colon \Ecal^1_{\td}(\RR^3  ,\gfrak ) \to \coker(l_1)$ we simply take the quotient map. In terms of these choices, the only non-zero component of the contracting homotopy $h$ turns out to be 
\begin{equation}
h^{(2)} = G^{\uW} \colon \Ecal^1_{\td}(\RR^3  ,\gfrak ) \longrightarrow \Ecal^1_{\td}(\RR^3  ,\gfrak ). 
\end{equation}
Having this, the quasi-ismorphism between $H^{\sbullet}(\Lfrak)$ and $\Lfrak$ is readily determined by maps $f_{n} \colon H^{\sbullet}(\Lfrak)^{\otimes n} \to \Lfrak$ which are constructed recursively using $h$, whereas the higher order brackets $l'_n \colon H^{\sbullet}(\Lfrak)^{\otimes n} \to H^{\sbullet}(\Lfrak)$ are constructed recursively using $p$. We shall not reproduce here the explicit expressions, but instead refer the reader to the Appendix~A of Ref.~\cite{Macrelli:2019afx} (see also Ref.~\cite{Jurco:2018sby}). 

We shall now obtain the perturbiner expansion for the non-abelian Navier-Stokes equation by using the minimal $L_{\infty}$-structure on $H^{\sbullet}(\Lfrak)$. To reach that goal, we consider a Maurer-Cartan element $u' \in H^{1}(\Lfrak) = \ker(l_1)$ with components of the form
\begin{equation}\label{eq:4.5}
u'_{i}(\xbf,t) = \sum_{p \geq 1} \ucal_{i p} \ue^{\ui(\kbf_{p} \cdot \xbf + \omega_p t)} T_{a_p},
\end{equation}
where $\omega_p$ is related to $\kbf_p$ through the dispersion relation $\ui \omega_p + \nu \kbf_p^2 = 0$. This, of course, is the simplest multiparticle solution to the linearised equation. We then define the colour-stripped perturbiner expansion to be the Maurer-Cartan element $u$ of $\Lfrak$ given by the formula
\begin{equation}\label{eq:4.6}
u = \sum_{n \geq 1} \frac{1}{n!} f_{n}(u',\dots,u'). 
\end{equation}
The task is to work out the components $u_{i}(\xbf,t)$ of $u$. Here we may borrow from the analysis carried out in Ref.~\cite{Lopez-Arcos:2019hvg}. To start off, a detailed calculation gives the term $f_{n}(u',\dots,u')$ in Eq.~\eqref{eq:4.6} as
\begin{equation}
f_{n}(u',\dots,u') = - \tfrac{1}{2} \sum_{k = 1}^{n-1} \binom{n}{k} G^{\uW} \left( \lbbar f_{k}(u',\dots,u'), f_{n-k}(u',\dots,u')\rbbar\right).
\end{equation}
Next, using mathematical induction and taking note of Eqs.~\eqref{eq:2.5} and ~\eqref{eq:4.1}, the components of the above result in the form
\begin{equation}\label{eq:4.8}
f_{n}(u',\dots,u')_{i} = n! \sum_{P \in \Wcal_n} \ucal_{i P} \ue^{\ui (\kbf_P \cdot \xbf + \omega_P t)} T_{a_P},
\end{equation}
where the coefficients $\ucal_{i P}$ are determined from the recursion relations
\begin{equation}\label{eq:4.9}
\ucal_{i P} = \frac{\lambda}{\ui \omega_P + \nu \kbf_P^2 }\sum_{P = Q R}  \big\{ (\ucal_Q^{\sharp}  \cdot \kbf_R) \ucal_{i R} - (\ucal_R^{\sharp} \cdot \kbf_Q) \ucal_{i Q}\big\}.
\end{equation}
Here the notation $\sum_{P=QR}$ instructs to sum over deconcatenations of the word $P$ into non-empty words $Q$ and $R$. In line with the terminology used in Ref.~\cite{Mizera:2018jbh}, we refer to the coefficients $\ucal_{i P}$ as the colour-stripped Berends-Giele currents for $u$. Inserting Eq.~\eqref{eq:4.8} back in Eq.~\eqref{eq:4.6} gives then the components of the perturbiner expansion as
\begin{equation}\label{eq:4.10}
u_{i}(\xbf,t) = \sum_{n \geq 1} \sum_{P \in \Wcal_n} \ucal_{i P} \ue^{\ui (\kbf_P \cdot \xbf + \omega_P t)} T_{a_P}.
\end{equation}
So our conclusion is that the recursion relations for the perturbiner coefficients are preset by the recursion relations for the $L_{\infty}$-quasi-isomorphism from $H^{\sbullet}(\Lfrak)$ onto $\Lfrak$.

It is possible to put Eq.~\eqref{eq:4.9} into a somewhat more pliable form. In the first place, we can eliminate the $\omega_P$ from the denominator of the right-hand side of Eq.~\eqref{eq:4.9} by using the dispersion relation for each of the terms entering in Eq.~\eqref{eq:4.5}, and find that such denominator is given by 
\begin{equation}
s_P = 2 \nu \sum_{\{p,q\} \subset P} \kbf_p \cdot \kbf_q.
\end{equation}
This expression may be considered as the nonrelativistic analogue of the Mandelstam variables.  In the second place, it is very convenient to invoke a combinatorial gadget, termed the ``binary tree map'' in Ref.~\cite{Bridges:2019siz}, that allows us to keep track of the bracketed words obtained by iterated recursion of Eq.~\eqref{eq:4.9}. We shall here refer to it as the colour-stripped Berends-Giele map. It is defined as the map $b_{\cs}$ acting on all words and determined recursively by
\begin{align}
\begin{split}
b_{\cs}(p) &= p, \\
b_{\cs}(P) &= \frac{1}{s_P} \sum_{P = QR} [b_{\cs}(Q),b_{\cs}(R)]. 
\end{split}
\end{align}
In the third place, as a matter of notation, for an arbitrary labelled object $U_P$, we bring the definition from Ref.~\cite{Bridges:2019siz} for the replacement of words by such object as
\begin{equation}
\llbracket U \rrbracket \circ P = U_{P}. 
\end{equation}
And in the fourth and last place, for every pair of bracketed words $\ell[P]$ and $\ell[Q]$, we recursively set
\begin{equation}\label{eq:4.14}
\varepsilon_{i[\ell[P],\ell[Q]]} = (\varepsilon_{\ell[P]}^{\sharp}  \cdot \kbf_Q) \varepsilon_{i \ell[Q]} - (\varepsilon_{\ell[Q]}^{\sharp} \cdot \kbf_P) \varepsilon_{i \ell[P]},
\end{equation}
with the agreement that $\varepsilon_{ip} = \ucal_{ip}$. Later we shall see that the labelled objects $\varepsilon_{i \ell[P]}$ are to be identified with kinematic numerators satisfying generalised Jacobi identities.  With all this in mind, we can rewrite the recursion relation in Eq.~\eqref{eq:4.9} in the form
\begin{equation}\label{eq:4.15}
\ucal_{i P} = \lambda \llbracket \varepsilon_{i} \rrbracket \circ b_{\cs}(P). 
\end{equation}
We have then a clear-cut way in which the colour-stripped Berends-Giele map $b_{\cs}$ provides a purely combinatorial realisation of the colour-stripped Berends-Giele current $\ucal_{i P}$. From this, in particular, follows the shuffle constraint $\ucal_{i P \shuffle Q} = 0$, which can be equivalently stated as $\ucal_{i P p Q} =(-1)^{\lvert P\rvert} \ucal_{i p(\bar{P} \shuffle Q)}$. This is seen at once by appealing to the general abstract argument presented in Appendix~C.1 of Ref.~\cite{Bridges:2019siz}. 


\section{Colour-dressed multiparticle solution to the non-abelian Navier-Stokes equation}\label{sec:5}
We are now concerned with the determination of a multiparticle solution to the non-abelian Navier-Stokes equations in the form of a colour-dressed perturbiner expansion. All conventions, notation and  terminology introduced in the previous section remain in force. 

The treatment of colour-dressed perturbiner expansions can be carried out along lines parallel to the treatment employed for colour-stripped perturbiner expansions and is in many respects simpler (see, for example, Ref.~\cite{Gomez:2020vat}). We start with a Maurer-Cartan element $u \in H^{\sbullet}(\Lfrak) = \ker(l_1)$ which has components
\begin{equation}
u'^{a}_{i}(\xbf,t) = \sum_{p \geq 1} \ucal^{a}_{i p} \ue^{\ui(\kbf_p \cdot \xbf + \omega_p t)}, 
\end{equation}
with $\omega_p$ and $\kbf_p$ obeying the dispersion relation $\ui \omega_p + \nu \kbf_p^2 = 0$. We next define the colour-dressed perturbiner expansion $u \in \Lfrak$ by means of the same formula as Eq.~\eqref{eq:4.6}. Again our basic problem will be to work out the components $u^{a}_{i}(\xbf,t)$ of $u$. This calculation proceeds exactly as before, using mathematical induction. We obtain
\begin{equation}\label{eq:5.2}
u^{a}_{i}(\xbf,t) = \sum_{n \geq 1} \sum_{P \in \OWcal_n} \ucal^{a}_{i P} \ue^{\ui (\kbf_P \cdot \xbf + \omega_P t)},
\end{equation}
where the coefficients $\ucal^{a}_{i P}$ are determined from the recursion relations
\begin{equation}\label{eq:5.3}
\ucal^{a}_{i P} = \frac{\lambda}{s_P} \sum_{P = Q \cup R}  \tfrac{1}{2} \tilde{f}_{bc}^{\phantom{bc}a}\big\{ (\ucal_Q^{b \sharp}  \cdot \kbf_R) \ucal^{c}_{i R} - (\ucal_R^{c \sharp} \cdot \kbf_Q) \ucal^{b}_{i Q}\big\}.
\end{equation}
Here the notation $P=Q \cup R$ instructs to distribute the letters of the ordered words $P$ into non-empty ordered words $Q$ and $R$. We have also introduced, for later notational convenience, the combination  $ \tilde{f}_{bc}^{\phantom{bc}a} = - 2 \ui  f_{bc}^{\phantom{bc}a}$. From now on the coefficients $\ucal^{a}_{i P}$ will be referred to as the colour-dressed Berends-Giele currents for $u$. It is, perhaps, worth mentioning that, in terms of the $\ucal^{a}_{i P}$, the solenoidal condition \eqref{eq:2.13a} translates into the transversality condition $\ucal^{a \sharp}_{P} \cdot \kbf_P = 0$.

We shall find the colour-dressed version of the perturbiner expansion extremely useful in our later work. In fact, we shall see that from the colour-dressed Berends-Giele currents we can extract kinematic numerators that naturally satisfy the generalised Jacobi identities of a nested commutator, an attribute that can be reconciled with a manifestation of colour-kinematics duality. 


\section{Factorisation of the non-abelian Navier-Stokes strict $L_{\infty}$-algebra}\label{sec:6}
In this section the problem of factoring the non-abelian Navier-Stokes strict $L_{\infty}$-algebra $\Lfrak$ is considered. More precisely, we are interested in establishing a factorisation of $\Lfrak$ into three parts: a colour part, a kinematic part and a strict $L_{\infty}$-algebra which fully describes the interaction of the perturbiner coefficients. This will be accomplished building upon the ideas and constructions of Ref.~\cite{Borsten:2021hua}. We shall therefore use the notation and terminology used there.

There are two steps in defining the required factorisation of the strict $L_{\infty}$-algebra $\Lfrak$. The first step amounts to showing that $\Lfrak$ admits a factorisation into a colour or gauge Lie algebra and a kinematical strict $C_{\infty}$-algebra $\Cfrak$. We start with the definition of the latter. As a cochain complex, it is given by
$$
\Omega^1_{\td}(\RR^3)[-1] \xrightarrow{\frac{\partial}{\partial t}+ \nu \delta \ud} \Omega^1_{\td}(\RR^3)[-2].
$$
In other words, the differential $m_1$ is identified with the operator $\frac{\partial}{\partial t}+ \nu \delta \ud$ acting on $\Omega^1_{\td}(\RR^3)$. The multiplication $m_2 \colon \Omega^1_{\td}(\RR^3)[-1]^{\otimes 2} \to \Omega^1_{\td}(\RR^3)[-2]$ is defined by taking
\begin{equation}\label{eq:6.1}
m_2(u,v) = \lambda [u,v].
\end{equation}
Clearly, this operation is graded commutative. Moreover, for degree reasons, the Leibniz rule is trivially satisfied. In this way, the graded vector space $\Cfrak = \Omega^1_{\td}(\RR^3)[-1] \oplus \Omega^1_{\td}(\RR^3)[-2]$ acquires the structure of a strict $C_{\infty}$-algebra. We claim that this strict $C_{\infty}$-algebra allows for a factorisation
\begin{equation}\label{eq:6.2}
\Lfrak = \gfrak \otimes \Cfrak,
\end{equation}
where the colour Lie algebra $\gfrak$ is concentrated in degree zero. To verify this, identify $\Omega^1_{\td}(\RR^{3},\gfrak)$ with  the tensor product $\gfrak \otimes \Omega^1_{\td}(\RR^{3})$ in the usual way, so that, we may write an element $u \in  \Omega^1_{\td}(\RR^{3},\gfrak)$ as $u = T_{a} \otimes u^{a}$ with $u^{a} \in \Omega^1_{\td}(\RR^{3})$. Then, recalling the definition of the differentials $l_1$ and $m_1$, we see that
\begin{align*}
l_1 (T_{a}\otimes u^{a}) = T_{a} \otimes \left( \frac{\partial}{\partial t}+ \nu \delta \ud\right) u^{a} = T_{a} \otimes m_1(u^{a}).
\end{align*}
Also, from Eqs.~\eqref{eq:3.1} and \eqref{eq:6.1}, we find that
\begin{align*}
l_2(T_{a} \otimes u^{a}, T_{b} \otimes v^{b}) &= \llbracket T_{a} \otimes u^{a}, T_{b} \otimes v^{b} \rrbracket =  \lambda f_{ab}^{\phantom{ab}c}T_{c} \otimes [u^{a},v^{b}] = [T_{a},T_{b}] \otimes m_2 (u^{a},v^{b}).
\end{align*}
These last two equalities show that  $l_1 = \id \otimes m_1$ and $l_2 = [,] \otimes m_2$, and consequently the claim holds true. We remark in passing that the factorisation \eqref{eq:6.2} corresponds to colour-stripping the strict $L_{\infty}$-algebra $\Lfrak$. 

In order to make further progress it is necessary to adjust the definition of the strict $C_{\infty}$-algebra $\Cfrak$ so as to include perturbiner expansions. Following the notation of Section~\ref{sec:3}, we denote by $\Ecal^{0}_{\td}(\RR^3)$ the space of time-dependent formal series of the form
\begin{equation}
\varphi(\xbf,t) = \sum_{n \geq 1} \sum_{P \in \OWcal_n} \varphi_P \ue^{\ui (\kbf_P \cdot \xbf + \omega_P t)},
\end{equation}
and by $\Ecal^{1}_{\td}(\RR^3)$ the space of time-dependent $1$-forms on $\RR^3$ with coefficients on $\Ecal^{0}_{\td}(\RR^3)$. In line with the preceding discussion, we refer to the elements of $\Ecal^{1}_{\td}(\RR^3)$ as colour-stripped perturbiner ansatzs. We must also say a bit about how to extend the operator $\frac{\partial}{\partial t}+ \nu \delta \ud$ to act on the space $\Ecal^{1}_{\td}(\RR^3)$. This is done in the standard way:~we first extend the definition of $\frac{\partial}{\partial t}+ \nu \delta \ud$ to the space $\Ecal^{0}_{\td}(\RR^3)$ by putting
\begin{equation}
\left( \frac{\partial}{\partial t}+ \nu \delta \ud\right)\varphi(\xbf,t)  =  \sum_{n \geq 1} \sum_{P \in \OWcal_n}(\ui \omega_P + \nu \kbf_P^2) \varphi_P \ue^{\ui (\kbf_P \cdot \xbf + \omega_P t)},
\end{equation}
and then extend $\frac{\partial}{\partial t}+ \nu \delta \ud$ uniquely to $\Ecal^{1}_{\td}(\RR^3)$ by linearity. 

Equipped with all this information, the cochain complex underlying the strict $C_{\infty}$-algebra $\Cfrak$ controlling the perturbiner expansion is
$$
\Ecal^1_{\td}(\RR^3)[-1] \xrightarrow{\frac{\partial}{\partial t}+ \nu \delta \ud} \Ecal^1_{\td}(\RR^3)[-2].
$$
Hence, the differential $m_1$ is simply the operator $\frac{\partial}{\partial t}+ \nu \delta \ud$ acting on $\Ecal^{1}_{\td}(\RR^3)$. Regarding the multiplication $m_2 \colon \Ecal^1_{\td}(\RR^3)[-1]^{\otimes 2} \to \Ecal^1_{\td}(\RR^3)[-2]$, this is determined by the exact same formula \eqref{eq:6.1} as above. Using this prescription, and recalling the definition of the strict $L_{\infty}$-algebra $\Lfrak$ that results in dealing with colour-dressed perturbiner expansions, we deduce that the factorisation \eqref{eq:6.2} remains valid.

With this background in mind, the second step consists of showing that the strict $C_{\infty}$-algebra $\Cfrak$ admits a further factorisation which strips off the kinematical factor. For this purpose, we use the notion of twisted tensor product of homotopy algebras introduced in Ref.~\cite{Borsten:2021hua}. We proceed to give some of the necessary definitions. 

Let $\Kin = (\RR^{3 })^{*}$ be the dual space of covectors in $\RR^3$, which we regard as being a graded vector space sitting in degree zero. We continue to write $e^{i}$ for the natural basis of $\Kin$ relative to the rectangular coordinates of $\RR^3$. We also consider the strict $L_{\infty}$-algebra $\Scal$ built from the cochain complex
$$
\Ecal^0_{\td}(\RR^3)[-1] \xrightarrow{\frac{\partial}{\partial t}+ \nu \delta \ud} \Ecal^0_{\td}(\RR^3)[-2],
$$
whose differential we write as $\mu_1$ and whose bracket $\mu_{2} \colon \Ecal^0_{\td}(\RR^3)[-1]^{\otimes 2} \to \Ecal^0_{\td}(\RR^3)[-2]$ is defined as follows:~for any two elements $\varphi(\xbf,t), \psi(\xbf,t) \in \Ecal^0_{\td}(\RR^3)$ we set
\begin{equation}\label{eq:6.5}
\mu_2(\varphi(\xbf,t),\psi(\xbf,t)) =  \sum_{n\geq 1} \sum_{P \in \OWcal_n}  \Bigg(  \sum_{P= Q \cup R}    \lambda  \varphi_{Q}\psi_{R}\Bigg)\ue^{\ui (\kbf_P \cdot \xbf + \omega_P t)},
\end{equation}
where $\varphi_P$ and $\psi_Q$ are the coefficients in the expansions of $\varphi(\xbf,t)$ and $\psi(\xbf,t)$ as formal series. We are interested here in computing the twisted tensor product between $\Kin$ and $\Scal$. To do so, we need to define a twist datum $\tau =(\tau_1,\tau_2)$ consisting of a pair of maps $\tau_1 \colon \Kin \to \Kin \otimes \End (\Scal)$ and $\tau_2 \colon \Kin \otimes \Kin \to \Kin \otimes \End (\Scal) \otimes \End (\Scal)$. We put
\begin{equation}\label{eq:6.6}
\tau_1 (e^{i}) = e^{i} \otimes \id
\end{equation}
and
\begin{equation}\label{eq:6.7}
\tau_2 (e^{i} \otimes e^{j}) =   e^{j} \otimes \id \otimes \partial^{i} - e^{i} \otimes \partial^{j} \otimes \id.
\end{equation}
Using the above formulae for $\tau_1$ and $\tau_2$ as well as the prescription of Ref.~\cite{Borsten:2021hua}, it is a fact that the tensor product $\Kin \otimes \Scal$ carries the structure of a strict $C_{\infty}$-algebra with differential 
\begin{equation}\label{eq:6.8}
m_1^{\tau} (e^{i} \otimes u_{i}(\xbf,t)) = e^{i} \otimes \mu_1 (u_{i}(\xbf,t))
\end{equation}
and multiplication
\begin{align}\label{eq:6.9}
\begin{split}
&m_2^{\tau} (e^{i} \otimes u_{i}(\xbf,t), e^{j} \otimes v_{j}(\xbf,t)) \\
& \qquad\qquad =  e^{j} \otimes \mu_2(u_{i}(\xbf,t), \partial^{i} v_{j}(\xbf,t)) - e^{i} \otimes \mu_2(\partial^{j} u_{i}(\xbf,t), v_{j}(\xbf,t)).
\end{split}
\end{align}
This strict $C_{\infty}$-algebra is defined to be the twisted tensor product of $\Kin$ and $\Scal$ and is denoted by $\Kin \otimes_{\tau} \Scal$. 

Now to the point. We claim that the strict $C_{\infty}$-algebra $\Cfrak$ factorises as
\begin{equation}\label{eq:6.10}
\Cfrak = \Kin \otimes_{\tau} \Scal.
\end{equation}
To substantiate the claim, we identify $\Ecal^1_{\td}(\RR^3)$ with the tensor product $(\RR^{3 })^{*}\otimes \Ecal^0_{\td}(\RR^3)$ so that we may write any element $u \in \Ecal^1_{\td}(\RR^3)$ as $u = e^{i} \otimes u_{i}(\xbf,t)$ with $u_{i}(\xbf,t) \in \Ecal^0_{\td}(\RR^3)$. Under this identification, it is clear that $\Cfrak = \Kin \otimes \Scal$ as graded vector spaces. Thus, to show \eqref{eq:6.10}, we only need to verify that $m_1 = m_1^{\tau}$ and $m_2 = m_2^{\tau}$. That the first equality holds follows immediately from the definitions of $m_1$ and $m_1^{\tau}$. In fact, both are given by the formula \eqref{eq:6.8}. For the second equality, we simply calculate both sides separately. So fix two elements $e^{i} \otimes u_{i}(\xbf,t)$ and $e^{j} \otimes v_{j}(\xbf,t)$ of $\Ecal^1_{\td}(\RR^3)$. Then Eq.~\eqref{eq:6.1} may be expressed more explicitly as
\begin{equation}
m_2(e^{i} \otimes u_{i}(\xbf,t),e^{j} \otimes v_{j}(\xbf,t)) =  e^{i} \otimes  \lambda \big\{ u_{j}(\xbf,t) \partial^{j} v_{i}(\xbf,t) - v_{j}(\xbf,t) \partial^{j} u_{i}(\xbf,t) \big\}.
\end{equation}
If we further write $\ucal_{i P}$ and $\vcal_{jQ}$ for the coefficients in the expansions of $u_{i}(\xbf,t)$ and $v_{j}(\xbf,t)$ as formal series, this is
\begin{align}\label{eq:6.12}
\begin{split}
&m_2(e^{i} \otimes u_{i}(\xbf,t),e^{j} \otimes v_{j}(\xbf,t)) \\
&\qquad \qquad = e^{i} \otimes \sum_{n \geq 1} \sum_{P \in \OWcal_n}  \Bigg(\sum_{P = Q \cup R}  \ui \lambda \big\{ (\ucal_{Q}^{\sharp} \cdot \kbf_{R}) \vcal_{i R} - (\vcal_{R}^{\sharp} \cdot \kbf_{Q}) \ucal_{i Q} \big\} \Bigg) \ue^{\ui(\kbf_P \cdot \xbf + \omega_P t)}.
\end{split}
\end{align}
On the other hand, using Eq.~\eqref{eq:6.5} gives
\begin{equation}\label{eq:6.13}
\mu_2(u_{i}(\xbf,t), \partial^{i} v_{j}(\xbf,t)) = \sum_{n \geq 1} \sum_{P \in \OWcal_n} \Bigg( \sum_{P = Q \cup R} \ui \lambda (\ucal_{Q}^{\sharp} \cdot \kbf_{R}) \vcal_{j R} \Bigg)  \ue^{\ui(\kbf_P \cdot \xbf + \omega_P t)}.
\end{equation}
and
\begin{equation}\label{eq:6.14}
\mu_2(\partial^{j} u_{i}(\xbf,t), v_{j}(\xbf,t)) =  \sum_{n \geq 1} \sum_{P \in \OWcal_n} \Bigg( \sum_{P = Q \cup R} \ui  \lambda (\vcal_{R}^{\sharp} \cdot \kbf_{Q}) \ucal_{i Q} \Bigg)  \ue^{\ui(\kbf_P \cdot \xbf + \omega_P t)}
\end{equation}
Inserting Eqs.~\eqref{eq:6.13} and \eqref{eq:6.14} into Eq.~\eqref{eq:6.9} we find that $m_2^{\tau}(e^{i} \otimes u_{i}(\xbf,t),e^{j} \otimes v_{j}(\xbf,t))$ is also equal to the right-hand side of Eq.~\eqref{eq:6.12}, which proves what  we set  out to show.

To summarise, we have shown that the strict $L_{\infty}$-algebra $\Lfrak$ encoding the colour-dressed perturbiner expansion for the non-abelian Navier-Stokes equation factorises as
\begin{equation}\label{eq:6.15}
\Lfrak = \gfrak \otimes (\Kin \otimes_{\tau} \Scal).
\end{equation}
In terms of this factorisation, the double copy prescription becomes quite readily apparent. First, however, we need to verify that the factorisation is compatible with colour-kinematics duality. We shall do so in the following section.


\section{Colour-kinematics duality}\label{sec:7}
In this section we investigate how colour-kinematics duality can be made manifest for the colour-dressed Berends-Giele currents produced by the non-abelian Navier-Stokes equation. Specifically, we identify an infinite-dimensional Lie algebra which determines the kinematic numerators of the colour-dressed perturbiner coefficients satisfying the same generalised Jacobi identities as their colour factors. This means, among other things, that the latter Lie algebra is dual to the colour Lie algebra $\gfrak$, in the sense suggested by the work of Bern, Carrasco and Johansson~\cite{Bern:2008qj}. 

To begin with it is of course evident that we can always decompose the colour and kinematic degrees of freedom of the single index colour-dressed Berends-Giele currents $\ucal_{ip}^{a}$ by writting
\begin{equation}\label{eq:7.1}
\ucal_{ip}^{a} = \delta^{a}_{\phantom{a}a_p} \varepsilon_{i p}.
\end{equation}
Here the $\varepsilon_{i p}$ may be regarded as the components of a covector $\varepsilon_{p}$ in $\RR^3$. That said, let us consider the infinite-dimensional Lie algebra $\gfrak'$ that is spanned by the $\varepsilon_p$ and whose Lie bracket is defined by
\begin{equation}\label{eq:7.2}
[\varepsilon_{p},\varepsilon_{q}] =  (\varepsilon_{p}^{\sharp} \cdot \kbf_{q}) \varepsilon_{q} - (\varepsilon_{q}^{\sharp} \cdot \kbf_{p}) \varepsilon_{p}.
\end{equation}
By treating the infinite integer index $p$ as the ``Fourier transform'' of a continuous variable in $\RR^3$, this  algebra may be recognised as the Lie algebra of infinitesimal spatial diffeomorphisms. This is to be contrasted with the discussion made in the last section of Ref.~\cite{Cheung:2020djz}. 

In what follows, we will make believe that the Lie algebras $\gfrak$ and $\gfrak'$ are reciprocal in the sense of allowing the duality between color and kinematics, also referred to as color-kinematics duality. To simplify the expressions that occur in the calculation, for each bracketed word $\ell[P]= \ell[p_1 p_2 \cdots p_n]$ of length $n$, we employ the notation $c^{a}_{P}$ to indicate the product of colour factors determined by
\begin{equation}\label{eq:7.3}
c^{a}_{\ell[P]} =  \tilde{f}_{a_{p_1} a_{p_2}}{}^{b}\tilde{f}_{b a_{p_3}}{}^{c} \cdots \tilde{f}_{d a_{p_{n-1}}}{}^{e} \tilde{f}_{e a_{p_{n}}}{}^{a},
\end{equation}
with the understanding that $c^{a}_{p} = \delta^{a}_{\phantom{a}a_{p}}$. We also set
\begin{equation}\label{eq:7.4}
c^{a}_{[\ell[P],\ell[Q]]} = \tilde{f}_{bc}^{\phantom{bc}a} c^{b}_{\ell[P]} c^{c}_{\ell[Q]}
\end{equation}
for every pair of bracketed words $\ell[P]$ and $\ell[Q]$. Using this notation we shall proceed to write down explicitly the Berends-Giele currents $u^{a}_{i P}$ by making direct reference to the components $\varepsilon_{i p}$ of the generators of $\gfrak'$. By way of preparation, we first take $P = 12$ in Eq.~\eqref{eq:5.3}. In this case, the possible ways of distributing the letters are $(Q,R) =(1,2), (2,1)$. Then, using Eqs.~\eqref{eq:7.1}--\eqref{eq:7.4}, we find that the colour-dressed Berends-Giele current $u^{a}_{i 12}$ acquires the form  
\begin{equation}\label{eq:7.5}
\ucal_{i 12}^{a} = \lambda \Bigg(\frac{c^{a}_{[1,2]} \varepsilon_{i [1,2]}}{s_{12}} \Bigg),
\end{equation}
where we have introduced, for convenience, the kinematic numerator $\varepsilon_{i [1,2]} = [\varepsilon_1, \varepsilon_2]_{i}$. It follows immediately that the $\varepsilon_{i [1,2]}$ satisfy the same antisymmetry properties under interchange of $1$ and $2$ as the colour factor $c^{a}_{[1,2]}$. Next take $P=123$ in Eq.~\eqref{eq:5.3}. In this case, the possible ways of distributing the letters are $(Q,R) =(12,3), (13,2),(23,1),(1,23),(2,13),(3,12)$. Therefore, after a straightforward calculation making use of Eqs.~\eqref{eq:7.1}--\eqref{eq:7.5}, we obtain for the  colour-dressed Berends-Giele current $\ucal_{i 123}^{a}$ the formula
\begin{align}\label{eq:7.6}
\ucal_{i 123}^{a} = \lambda^2 \Bigg( \frac{c^{a}_{[[1,2],3]} \varepsilon_{i [[1,2],3]}}{s_{12} s_{123}} + \frac{c^{a}_{[[1,3],2]} \varepsilon_{i [[1,3],2]}}{s_{13} s_{123}} + \frac{c^{a}_{[[2,3],1]} \varepsilon_{i [[2,3],1]}}{s_{23} s_{123}}  \Bigg),
\end{align}
where now the kinematic numerators are $\varepsilon_{i [[1,2],3]} = [[\varepsilon_{1},\varepsilon_{2}],\varepsilon_{3}]_{i}$, $\varepsilon_{i [[1,3],2]} = [[\varepsilon_{1},\varepsilon_{3}],\varepsilon_{2}]_{i}$ and $\varepsilon_{i [[2,3],1]} = [[\varepsilon_{2},\varepsilon_{3}],\varepsilon_{1}]_{i}$. Notice in particular that the last two terms in the parenthesis in  Eq.~\eqref{eq:6.6} are obtained from the first by permuting the indices $1$, $2$ and $3$ cyclically. Furthermore, it may be emphasised again that the Jacobi identity, which requires $\varepsilon_{i [[1,2],3]}$ to vanish when antisymmetrised on $1$, $2$ and $3$, mirrors the Jacobi identity satisfied by the colour factor $c^{a}_{[[1,2],3]}$. This matter can perhaps be made a little plainer if we subsequently take $P = 1234$ in Eq.~\eqref{eq:5.3}. In this case, the possible ways of distributing the letters that contribute to the sum are $(Q,R) = (123,4)$, $(124,3)$, $(134,2)$, $(234,1)$, $(12,34)$, $(13,24)$, $(23,14)$, $(1,234)$, $(2,134)$, $(3,124)$, $(4,123)$. By analogy with the calculation leading to Eq.~\eqref{eq:7.6}, it is not difficult, though perhaps a little tedious, to verify that the colour-dressed Berends-Giele current $\ucal_{i 1234}^{a}$ may be represented in the form 
\begin{align}\label{eq:7.7}
\begin{split}
\ucal_{i 1234}^{a} =\lambda^3 \Bigg( & \frac{c^{a}_{[[[1,2],3],4]} \varepsilon_{i [[[1,2],3],4]}}{s_{12} s_{123} s_{1234}} + \frac{c^{a}_{[[[1,2],4],3]} \varepsilon_{i [[[1,2],4],3]}}{s_{12} s_{124} s_{1234}} +\frac{c^{a}_{[[[1,3],2],4]} \varepsilon_{i [[[1,3],2],4]}}{s_{13} s_{123} s_{1234}} \\
&+ \frac{c^{a}_{[[[1,3],4],2]} \varepsilon_{i [[[1,3],4],2]}}{s_{13} s_{134} s_{1234}} +\frac{c^{a}_{[[[1,4],2],3]} \varepsilon_{i [[[1,4],2],3]}}{s_{14} s_{124} s_{1234}} + \frac{c^{a}_{[[[1,4],3],2]} \varepsilon_{i [[[1,4],3],2]}}{s_{14} s_{134} s_{1234}} \\
&+\frac{c^{a}_{[[[2,3],1],4]} \varepsilon_{i [[[2,3],1],4]}}{s_{23} s_{123} s_{1234}} + \frac{c^{a}_{[[[2,3],4],1]} \varepsilon_{i [[[2,3],4],1]}}{s_{23} s_{234} s_{1234}} +\frac{c^{a}_{[[[2,4],1],3]} \varepsilon_{i [[[2,4],1],3]}}{s_{24} s_{124} s_{1234}} \\
& +\frac{c^{a}_{[[[2,4],3],1]} \varepsilon_{i [[[2,4],3],1]}}{s_{24} s_{234} s_{1234}} +\frac{c^{a}_{[[[3,4],1],2]} \varepsilon_{i [[[3,4],1],2]}}{s_{34} s_{134} s_{1234}} +\frac{c^{a}_{[[[3,4],2],1]} \varepsilon_{i [[[3,4],2],1]}}{s_{34} s_{234} s_{1234}} \\
&+\frac{c^{a}_{[[1,2],[3,4]]} \varepsilon_{i [[1,2],[3,4]]}}{s_{12} s_{34} s_{1234}} + \frac{c^{a}_{[[1,3],[2,4]]} \varepsilon_{i [[1,3],[2,4]]}}{s_{13} s_{24} s_{1234}} +\frac{c^{a}_{[[1,4],[2,3]]} \varepsilon_{i [[1,4],[2,3]]}}{s_{14} s_{23} s_{1234}} \Bigg),
\end{split}
\end{align}
where, as the notation implies, the kinematic numerators are $\varepsilon_{i [[[1,2],3],4]} = [[[\varepsilon_{1},\varepsilon_{2}],\varepsilon_{3}],\varepsilon_{4}]_{i}$, $\varepsilon_{i [[1,2],[3,4]]}= [[\varepsilon_{1},\varepsilon_{2}],[\varepsilon_{3},\varepsilon_{4}]]_{i}$, and so on. Thus we are once more led to conclude that these kinematic numerators share the same symmetry properties as the corresponding colour factors.

The above pattern continues as we keep increasing the length of the word $P$. To see it, we need yet some more notation. In the first place, for any bracketed word $\ell[P] = \ell[p_1p_2 \cdots p_n]$ of length $n$, the kinematic numerator $\varepsilon_{i\ell[P]}$ is defined to be
\begin{equation}\label{eq:7.8}
\varepsilon_{i \ell[P]} =  [[\dots[ [\varepsilon_{p_1},\varepsilon_{p_2}],\varepsilon_{p_3}], \dots], \varepsilon_{p_n}]_i.
\end{equation}
We further put
\begin{equation}
\varepsilon_{i [\ell[P],\ell[Q]]} = [\varepsilon_{\ell[P]},\varepsilon_{\ell[Q]}]_i
\end{equation}
for every pair of bracketed words $\ell[P]$ and $\ell[Q]$. One easily checks by employing Eq.~\eqref{eq:7.2} that this reproduces Eq.~\eqref{eq:4.14}. In the second place, we need to modify the colour-stripped Berends-Giele map by a colour-dressed version of it. To be more precise, here we consider the map $b_{\cd}$ acting on ordered words and determined recursively by
\begin{align}\label{eq:7.9}
\begin{split}
b_{\cd}(p) &= p, \\
b_{\cd}(P) &= \frac{1}{2s_P} \sum_{P = Q \cup R} [b_{\cd}(Q),b_{\cd}(R)]. 
\end{split}
\end{align}
Parenthetically it may be worth remarking that the factor of $2$ in the denominator on the right-hand side of the second formula in Eq.~\eqref{eq:6.9} can be dropped if we impose the condition that $\lvert  Q \rvert \geq \lvert  R \rvert$. And, in the third place, given two arbitrary labelled objects $U_P$ and $V_P$, we define the replacement of ordered words by the product of such objects as
\begin{equation}
\llbracket U \otimes V \rrbracket \circ P = U_{P} V_{P}. 
\end{equation}
With all the foregoing, it can be shown that the recursion relation in Eq.~\eqref{eq:5.3} is expressible in the form
\begin{equation}\label{eq:7.12}
\ucal^{a}_{i P} = \lambda^{\lvert P \rvert-1} \llbracket c^{a} \otimes \varepsilon_{i} \rrbracket \circ b_{\cd}(P).
\end{equation}
Taking note of Eq.~\eqref{eq:7.9}, this amounts to saying that the generalised Jacobi identities associated to the colour factors $c^{a}_{\ell[P]}$ are also obeyed by the kinematic numerators $\varepsilon_{i \ell[P]}$. Since the latter are, according to Eq.~\eqref{eq:7.7}, built out of structure constants of the infinite dimensional Lie algebra $\gfrak'$, we may conclude that $\gfrak'$ constitutes a particular realisation of the ``kinematic Lie algebra'' that underlies the duality between colour and kinematics.\footnote{It is worth noting that this algebra has been studied before in the context of colour-kinematics duality for the self-dual sector of Yang-Mills theory in Refs.~\cite{Monteiro:2011pc,Bjerrum-Bohr:2012kaa,Monteiro:2013rya,Fu:2016plh}, $3$-dimensional Chern-Simions theory in Ref.~\cite{Ben-Shahar:2021zww}, the non-linear sigma model in Ref.~\cite{Cheung:2021zvb}, and $10$-dimensional super Yang-Mills theory in Ref.~\cite{Ben-Shahar:2021doh}.} It should also be borne in mind that the ``factorisation'' of the colour-dressed Berends-Giele currents given in Eq.~\eqref{eq:7.12} is a manifestation of the factorisation \eqref{eq:6.2} of the strict $L_{\infty}$-algebra $\Lfrak$. 

The preceding discussion provides the specific justification for using the double copy prescription. In doing so, we shall appreciate more fully the power of the $L_{\infty}$-language at work for us.


\section{The double copy of the non-abelian Navier-Stokes equation}\label{sec:8}
It is our intention in this section to implement the double copy of the non-abelian Navier-Stokes equation. We shall begin by discussing the strict $L_{\infty}$-algebra which captures the dynamics described by such double copy. Following this we shall show that this strict $L_{\infty}$-algebra may be obtained directly from the ``homotopy double copy'' procedure outlined in Ref.~\cite{Borsten:2021hua}, which is implied by the factorisation of the non-abelian Navier-Stokes strict $L_{\infty}$-algebra $\Lfrak$ examined in Section~\ref{sec:6}. 

Let $\Omega^{1}_{\td}(\RR^3) \otimes_{\Omega^{0}_{\td}(\RR^3)} \Omega^{1}_{\td}(\RR^3)$ denote the tensor product of $\Omega^{1}_{\td}(\RR^3)$ with itself considered as a module over $\Omega^{0}_{\td}(\RR^3)$. We think of the elements of $\Omega^{1}_{\td}(\RR^3) \otimes_{\Omega^{0}_{\td}(\RR^3)} \Omega^{1}_{\td}(\RR^3)$ as time-dependent $1$-forms on $\RR^3$ with values in the Lie algebra $\Omega^{1}_{\td}(\RR^3)$. Thus, with reference to a basis $e^{\bar{i}}$ corresponding to a rectangular coordinate system $(x^{\bar{i}})$, an element $u \in \Omega^{1}_{\td}(\RR^3) \otimes_{\Omega^{0}_{\td}(\RR^3)} \Omega^{1}_{\td}(\RR^3)$ is decomposed as $u = e^{\bar{i}} \otimes u_{\bar{i}}$ with $u_{\bar{i}} \in \Omega^{1}_{\td}(\RR^3)$. We may therefore extend the operator $\frac{\partial}{\partial t} + \nu \delta \ud$ to act on $\Omega^{1}_{\td}(\RR^3) \otimes_{\Omega^{0}_{\td}(\RR^3)} \Omega^{1}_{\td}(\RR^3)$ by putting
\begin{equation}\label{eq:8.1}
\left( \frac{\partial}{\partial t} + \nu \delta \ud\right) u = e^{\bar{i}} \otimes \left( \frac{\partial u_{\bar{i}}}{\partial t} + \nu \delta \ud u_{\bar{i}}\right).
\end{equation}
Bearing this in mind, the cochain complex underlying the double copy strict $L_{\infty}$-algebra $\Lfrak^{\prime}$ is simply
$$
\Omega^{1}_{\td}(\RR^3) \otimes_{\Omega^{0}_{\td}(\RR^3)} \Omega^{1}_{\td}(\RR^3)[-1]  \xrightarrow{\frac{\partial}{\partial t}+ \nu \delta \ud} \Omega^{1}_{\td}(\RR^3) \otimes_{\Omega^{0}_{\td}(\RR^3)} \Omega^{1}_{\td}(\RR^3)[-2].
$$
Henceforth, as is customary, the symbol $l_1$ is used to represent the differential. To define the bracket, we first define a binary operation on $\Omega^{1}_{\td}(\RR^3) \otimes_{\Omega^{0}_{\td}(\RR^3)} \Omega^{1}_{\td}(\RR^3)$ as follows. Let $u \in \Omega^{1}_{\td}(\RR^3) \otimes_{\Omega^{0}_{\td}(\RR^3)} \Omega^{1}_{\td}(\RR^3)$ and let $u_{\bar{i}}$ the components of $u$ relative to the basis $e^{\bar{i}}$. We also write $u_{\bar{i} i}$ for the components of $u_{\bar{i}}$ relative to the given basis $e^{i}$. Since $x^{i}$ and $x^{\bar{i}}$ are independent variables the partial derivatives $\partial^{i}$ and $\partial^{\bar{i}}$ commute and hence it makes sense to define $\partial^{\bar{j}} u_{\bar{i}}$ as the time-dependent $1$-form on $\RR^3$ whose components relative to $e^{i}$ are $\partial^{\bar{j}} u_{\bar{i} i}$. With this understood, we define the binary operation by the formula
\begin{equation}\label{eq:8.2}
\lbbar u,v \rbbar = e^{\bar{i}} \otimes \frac{\kappa}{2} \big\{ [u_{\bar{j}}, \partial^{\bar{j}}v_{\bar{i}}] + [v_{\bar{j}}, \partial^{\bar{j}}u_{\bar{i}}] \big\},
\end{equation}
where $\kappa$ is a coupling constant. Written explicitly in components this formula has the form
\begin{equation}\label{eq:8.3}
\lbbar u,v \rbbar_{\bar{i} i} = \frac{\kappa}{2}(u_{\bar{j}j}  \partial^{\bar{j}} \partial^{j}  v_{\bar{i} i}  -  \partial^{\bar{j}} v_{\bar{i} j}\partial^{j} u_{\bar{j}i} + v_{\bar{j}j} \partial^{\bar{j}} \partial^{j} u_{\bar{i} i} -  \partial^{\bar{j}} u_{\bar{i} j}\partial^{j} v_{\bar{j}i}).
\end{equation}
The bracket $l_2 \colon (\Omega^{1}_{\td}(\RR^3) \otimes_{\Omega^{0}_{\td}(\RR^3)} \Omega^{1}_{\td}(\RR^3)[-1])^{\otimes 2}  \to \Omega^{1}_{\td}(\RR^3) \otimes_{\Omega^{0}_{\td}(\RR^3)} \Omega^{1}_{\td}(\RR^3)[-2]$ can now be obtained by simply setting
\begin{equation}\label{eq:8.4}
l_2 (u,v) = \lbbar u,v \rbbar. 
\end{equation}
As this is clearly skew-symmetric and trivially satisfies the graded Jacobi identity, the graded vector space $\Lfrak^{\prime} =(\Omega^{1}_{\td}(\RR^3) \otimes_{\Omega^{0}_{\td}(\RR^3)} \Omega^{1}_{\td}(\RR^3)[-1]) \oplus (\Omega^{1}_{\td}(\RR^3) \otimes_{\Omega^{0}_{\td}(\RR^3)} \Omega^{1}_{\td}(\RR^3)[-2])$ is indeed a strict $L_{\infty}$-algebra. 

With the help of the foregoing we may now obtain an expression for the field equation that governs the dynamics of the double copy. This is an extremely simple matter:~just set down the Maurer-Cartan equation associated to $\Lfrak^{\prime}$. From the definitions which we have given of the differential $l_1$ and the bracket $l_2$, the latter equation reads
\begin{equation}\label{eq:8.5}
\frac{\partial u}{\partial t} + \nu \delta \ud u + \tfrac{1}{2}  \lbbar u, u \rbbar = 0. 
\end{equation}
It should be specially noted that this equation is identical in form with the non-abelian Navier-Stokes equation~\eqref{eq:2.12b}. To emphasise this we have adopted the same symbol $\lbbar , \rbbar$ to designate the relevant binary operation. By using Eqs.~\eqref{eq:8.1} and~\eqref{eq:8.3}, we find the component form of Eq.~\eqref{eq:8.5} to be
\begin{equation}\label{eq:8.6}
\frac{\partial u_{\bar{i} i}}{\partial t} - \nu \Delta u_{\bar{i} i} + \frac{\kappa}{2}( u_{\bar{j}j}  \partial^{\bar{j}} \partial^{j} u_{\bar{i} i}  -  \partial^{\bar{j}} u_{\bar{i} j}\partial^{j} u_{\bar{j}i}  ) = 0.
\end{equation}
On account of the similarity between Eqs.~\eqref{eq:2.13b} and~\eqref{eq:8.6} we could speak of the latter as a tensor Navier-Stokes equation describing the dynamics of a bi-fluid velocity distribution with components $u_{\bar{i} i}$. The terminology is taken from Ref.~\cite{Cheung:2020djz} with a view towards our precise needs.

To complete our consideration of the double copy strict $L_{\infty}$-algebra $\Lfrak^{\prime}$ we must now readjust its definition to contend with perturbiner expansions. Here we return to the notation described in Section~\ref{sec:6}. Consider the tensor product $\Ecal^{1}_{\td}(\RR^3) \otimes_{\Ecal^{0}_{\td}(\RR^3)} \Ecal^{1}_{\td}(\RR^3)$ and expand once again its elements as $u = e^{\bar{i}} \otimes u_{\bar{i}}$ with $u_{\bar{i}} \in  \Ecal^{1}_{\td}(\RR^3)$. Of course, this means that each of the components $u_{\bar{i}i}(\xbf,t)$ of $u_{\bar{i}}$ with respect to the basis $e^{i}$ is expressible by means of a formal series of the form
\begin{equation}
u_{\bar{i}i}(\xbf,t) = \sum_{n \geq 1} \sum_{P \in \OWcal_n} \ucal_{\bar{i}i P} \ue^{\ui (\kbf_P \cdot \xbf + \omega_P t)}. 
\end{equation}
The cochain complex underlying the strict $L_{\infty}$-algebra $\Lfrak^{\prime}$ that encodes the perturbiner expansion for the tensor Navier-Stokes equation is thus
\begin{equation}\label{eq:8.8}
\Ecal^{1}_{\td}(\RR^3) \otimes_{\Ecal^{0}_{\td}(\RR^3)} \Ecal^{1}_{\td}(\RR^3)[-1]  \xrightarrow{\frac{\partial}{\partial t}+ \nu \delta \ud} \Ecal^{1}_{\td}(\RR^3) \otimes_{\Ecal^{0}_{\td}(\RR^3)} \Ecal^{1}_{\td}(\RR^3)[-2].
\end{equation}
The differential $l_1$ is therefore given simply by the obvious extension of Eq.~\eqref{eq:8.1}. As regards the bracket $l_2 \colon (\Ecal^{1}_{\td}(\RR^3) \otimes_{\Ecal^{0}_{\td}(\RR^3)} \Ecal^{1}_{\td}(\RR^3)[-1])^{\otimes 2} \to \Ecal^{1}_{\td}(\RR^3) \otimes_{\Ecal^{0}_{\td}(\RR^3)} \Ecal^{1}_{\td}(\RR^3)[-2]$, it has the same formula as that of Eq.~\eqref{eq:8.3}.

We go on now to cast the definition the double copy strict $L_{\infty}$-algebra $\Lfrak^{\prime}$ into a somewhat more elegant form by connecting it up with the homotopy double copy construction. The claim is that $\Lfrak^{\prime}$ can be obtained by replacing the colour factor $\gfrak$ in the factorisation \eqref{eq:6.15} of $\Lfrak$ with another copy of the ``twisted'' kinematic factor $\Kin$, while sending $\lambda$ to $\frac{\kappa}{2}$. Explicitly,
\begin{equation}\label{eq:8.9}
\Lfrak^{\prime}  = \Kin \otimes_{\tau} ( \Kin \otimes_{\tau} \Scal). 
\end{equation}
To establish this claim we must first specify the strict $L_{\infty}$-algebra structure on the twisted tensor product of the right hand side. To do this we note that an arbitrary element of  $\Kin \otimes_{\tau} ( \Kin \otimes_{\tau} \Scal)$ can be represented in the form $e^{\bar{i}} \otimes e^{i} \otimes u_{\bar{i}i}(\xbf,t)$ with $u_{\bar{i}i}(\xbf,t) \in \Ecal^{0}_{\td}(\RR^{3})$. Referring back to Eqs.~\eqref{eq:6.8} and~\eqref{eq:6.9} and the discussion in the ensuing paragraph, we consider $\Kin \otimes_{\tau} ( \Kin \otimes_{\tau} \Scal)$ as a strict $L_{\infty}$-algebra with differential 
\begin{align}\label{eq:8.10}
l_{1}^{\tau}(e^{\bar{i}} \otimes e^{i} \otimes u_{\bar{i}i}(\xbf,t)) = e^{\bar{i}} \otimes e^{i} \otimes \mu_1( u_{\bar{i}i}(\xbf,t))
\end{align}
and bracket
\begin{align}\label{eq:8.11}
\begin{split}
&l_{2}^{\tau}(e^{\bar{i}} \otimes e^{i} \otimes u_{\bar{i}i}(\xbf,t),e^{\bar{j}} \otimes e^{j} \otimes v_{\bar{j}j}(\xbf,t)) \\
& \qquad = e^{\bar{j}} \otimes e^{j} \otimes \mu_{2}(u_{\bar{i}i}(\xbf,t),  \partial^{\bar{i}} \partial^{i} v_{\bar{j}j}(\xbf,t) ) - e^{\bar{j}} \otimes e^{i} \otimes \mu_{2}(\partial^{j}u_{\bar{i}i}(\xbf,t), \partial^{\bar{i}} v_{\bar{j}j}(\xbf,t)) \\
& \qquad \quad \, + e^{\bar{i}} \otimes e^{i} \otimes \mu_{2}(\partial^{\bar{j}} \partial^{j} u_{\bar{i}i}(\xbf,t),v_{\bar{j}j}(\xbf,t)) -  e^{\bar{i}} \otimes e^{j} \otimes \mu_{2}(\partial^{\bar{j}} u_{\bar{i}i}(\xbf,t),\partial^{i} v_{\bar{j}j}(\xbf,t)).
\end{split}
\end{align}
Having clarified this point, let us return to the verification of \eqref{eq:8.9}. The first point to be noticed is that $\Ecal^{1}_{\td}(\RR^3) \otimes_{\Ecal^{0}_{\td}(\RR^3)} \Ecal^{1}_{\td}(\RR^3)$ can be naturally identified with $(\RR^{3 })^{*}\otimes (\RR^{3 })^{*} \otimes \Ecal^{0}_{\td}(\RR^3)$ so that every element of the former can be expanded in the form described above.  Evidently then, $\Lfrak^{\prime}  = \Kin \otimes ( \Kin \otimes \Scal)$ as graded vector spaces. Hence we are left to check that $l_1 = l_1^{\tau}$ and $l_2 = l_2^{\tau}$. The first equality is obvious since both $l_1$ and $l_1^{\tau}$ are determined by the same formula \eqref{eq:8.10}. For the second, let us take two elements $e^{\bar{i}} \otimes e^{i} \otimes u_{\bar{i}i}(\xbf,t)$ and $e^{\bar{j}} \otimes e^{j} \otimes v_{\bar{j}j}(\xbf,t)$ of $(\RR^{3 })^{*}\otimes (\RR^{3 })^{*} \otimes \Ecal^{0}_{\td}(\RR^3)$. By use of Eq.~\eqref{eq:8.3} we see that Eq.~\eqref{eq:8.4} may be rewritten as
\begin{align}
\begin{split}
&l_2(e^{\bar{i}} \otimes e^{i} \otimes u_{\bar{i}i}(\xbf,t),e^{\bar{j}} \otimes e^{j} \otimes v_{\bar{j}j}(\xbf,t)) \\
&\qquad  = e^{\bar{i}} \otimes e^{i} \otimes \frac{\kappa}{2} \big\{ u_{\bar{j}j}(\xbf,t)  \partial^{\bar{j}} \partial^{j}  v_{\bar{i} i}(\xbf,t) -  \partial^{\bar{j}} v_{\bar{i} j}(\xbf,t) \partial^{j} u_{\bar{j}i}(\xbf,t)\\
& \qquad \qquad \qquad \qquad\quad + v_{\bar{j}j}(\xbf,t) \partial^{\bar{j}} \partial^{j} u_{\bar{i} i}(\xbf,t) -  \partial^{\bar{j}} u_{\bar{i} j}(\xbf,t)\partial^{j} v_{\bar{j}i}(\xbf,t)\big\}.
\end{split}
\end{align}
Denoting the coefficients in the expansions of $u_{\bar{i}i}(\xbf,t)$ and $v_{\bar{j}j}(\xbf,t)$ by $\ucal_{\bar{i}i P}$ and $\vcal_{\bar{j}j Q}$, respectively, this becomes
\begin{align}\label{eq:8.13}
\begin{split}
&l_2(e^{\bar{i}} \otimes e^{i} \otimes u_{\bar{i}i}(\xbf,t),e^{\bar{j}} \otimes e^{j} \otimes v_{\bar{j}j}(\xbf,t)) \\
&\qquad  = e^{\bar{i}} \otimes e^{i} \otimes \sum_{n \geq 1} \sum_{P \in \OWcal_n}  \Bigg( \sum_{P = Q \cup  R}\frac{\kappa}{2} \big\{ k^{\bar{j}}_{R} \vcal_{\bar{i} j R} k^{j}_{Q} \ucal_{\bar{j}i Q}    - \ucal_{\bar{j}j Q}  k^{\bar{j}}_{R} k^{j}_{R}  \vcal_{\bar{i} i R} \\
& \qquad \qquad \qquad \qquad\qquad \qquad\qquad \quad\quad  + k^{\bar{j}}_{Q} \ucal_{\bar{i} j Q}k^{j}_{R} \vcal_{\bar{j}i R} - \vcal_{\bar{j}j R} k^{\bar{j}}_{Q} k^{j}_{Q} \ucal_{\bar{i} i Q} \big\} \Bigg)\ue^{\ui (\kbf_{P} \cdot \xbf + \omega_{P} t)}.
\end{split}
\end{align}
 At the same time, applying Eq.~\eqref{eq:6.5}, but with $\lambda$ replaced with $\frac{\kappa}{2}$, we obtain
 \begin{align}
 \begin{split}
  \mu_{2}(u_{\bar{i}i}(\xbf,t),  \partial^{\bar{i}} \partial^{i} v_{\bar{j}j}(\xbf,t) ) &= -\sum_{n \geq 1} \sum_{P \in \OWcal_n}  \Bigg( \sum_{P = Q \cup  R} \frac{\kappa}{2} \ucal_{\bar{i}i Q} k^{\bar{i}}_{R} k^{i}_{R} \vcal_{\bar{j} j R}  \Bigg) \ue^{\ui (\kbf_{P} \cdot \xbf + \omega_{P} t)},\\
 \mu_{2}(\partial^{j}u_{\bar{i}i}(\xbf,t), \partial^{\bar{i}} v_{\bar{j}j}(\xbf,t)) &= -\sum_{n \geq 1} \sum_{P \in \OWcal_n}  \Bigg( \sum_{P = Q \cup  R} \frac{\kappa}{2} k^{j}_{Q} \ucal_{\bar{i}i Q} k^{\bar{i}}_{R} \vcal_{\bar{j}j R}  \Bigg) \ue^{\ui (\kbf_{P} \cdot \xbf + \omega_{P} t)},\\
 \mu_{2}(\partial^{\bar{j}}\partial^{j} u_{\bar{i}i}(\xbf,t),v_{\bar{j}j}(\xbf,t)) &= -\sum_{n \geq 1} \sum_{P \in \OWcal_n}  \Bigg( \sum_{P = Q \cup  R} \frac{\kappa}{2} k^{\bar{j}}_{Q} k^{j}_{Q} \ucal_{\bar{i} i Q} \vcal_{\bar{j} j R} \Bigg) \ue^{\ui (\kbf_{P} \cdot \xbf + \omega_{P} t)}, \\
 \mu_{2}(\partial^{\bar{j}} u_{\bar{i}i}(\xbf,t),\partial^{i} v_{\bar{j}j}(\xbf,t)) &= -\sum_{n \geq 1} \sum_{P \in \OWcal_n}  \Bigg( \sum_{P = Q \cup  R} \frac{\kappa}{2} k^{\bar{j}}_{Q} \ucal_{\bar{i} i Q} k^{i}_{R} \vcal_{\bar{j}j R}  \Bigg) \ue^{\ui (\kbf_{P} \cdot \xbf +\omega_{P} t)}.
 \end{split}
 \end{align} 
 Substituting these expressions back into Eq.~\eqref{eq:8.11}, we find that $l_{2}^{\tau}(e^{\bar{i}} \otimes e^{i} \otimes u_{\bar{i}i}(\xbf,t),e^{\bar{j}} \otimes e^{j} \otimes v_{\bar{j}j}(\xbf,t))$ also equals the right-hand side of Eq.~\eqref{eq:8.13}, as was to be shown. 

To sum up, we could have avoided the guesswork of figuring out precisely what the double copy procedure is supposed to do to the non-abelian Navier-Stokes equation by appealing directly to the homotopy double copy recipe. In our opinion, this framework offers a more systematic and rigorous method of tackling the problem. 

\section{Multiparticle solution to the tensor Navier-Stokes equation}\label{sec:9}
We may now turn to the multiparticle solution to the tensor Navier-Stokes equation. Again we use what we have learned in the preceding section to write this solution in the form of a perturbiner expansion. 

We first describe the minimal $L_{\infty}$-structure on the cohomology $H^{\sbullet}(\Lfrak')$ of the double copy strict $L_{\infty}$-algebra $\Lfrak'$. Like before, all we need to do is to define a projection $p \colon \Lfrak' \to H^{\sbullet}(\Lfrak')$ and a contracting homotopy $h \colon \Lfrak' \to \Lfrak'$. As a preliminary remark, note that by virtue of the definition \eqref{eq:8.8}, the cochain complex underlying $H^{\sbullet}(\Lfrak')$ is
$$
\ker (l_1)[-1] \xrightarrow{\phantom{aa} 0 \phantom{aa}} \coker (l_1)[-2].
$$
This is formally identical to the one we encountered in Section~\ref{sec:4}. We thus proceed as we did there. To start, we extend the Wyld propagator $G^{\uW}$ so that we obtain a linear operator $G^{\uW} \colon \Ecal^{1}_{\td}(\RR^3) \otimes_{\Ecal^{0}_{\td}(\RR^3)} \Ecal^{1}_{\td}(\RR^3) \to \Ecal^{1}_{\td}(\RR^3) \otimes_{\Ecal^{0}_{\td}(\RR^3)} \Ecal^{1}_{\td}(\RR^3)$. This enables us to define the projection $p^{(1)}\colon \Ecal^{1}_{\td}(\RR^3) \otimes_{\Ecal^{0}_{\td}(\RR^3)} \Ecal^{1}_{\td}(\RR^3)  \to \ker (l_1)$ by the formula
\begin{equation}
p^{(1)} = \id_{\Ecal^{1}_{\td}(\RR^3) \otimes_{\Ecal^{0}_{\td}(\RR^3)} \Ecal^{1}_{\td}(\RR^3)} - G^{\uW}\circ l_1.
\end{equation}
The other projection $p^{(2)}\colon \Ecal^{1}_{\td}(\RR^3) \otimes_{\Ecal^{0}_{\td}(\RR^3)} \Ecal^{1}_{\td}(\RR^3) \to  \coker (l_1)$ we take simply as the quotient map. Lastly, it can be seen that the only non-zero component of the contracting homotopy $h$ is given by $h^{(2)} = G^{\uW}$. 

We are now ready to formulate the perturbiner expansion for the tensor Navier-Stokes equation. First of all, we pick a Maurer-Cartan element $u' \in H^{1}(\Lfrak') = \ker(l_1)$, for which we have 
\begin{equation}
u'_{\bar{i}i}(\xbf,t) = \sum_{p \geq 1} \ucal_{\bar{i}i p} \ue^{\ui (\kbf_p \cdot \xbf + \omega_p t)}. 
\end{equation}
Then we define the perturbiner expansion to be the Maurer-Cartan element $u$ in $\Lfrak'$ given by the formula
\begin{equation}
u = \sum_{n \geq 1} \frac{1}{n!} f_n(u',\dots,u').
\end{equation}
We want to calculate the components of $u$. Obviously the same argument that led to Eq.~\eqref{eq:5.2} applies here, and gives the components of $u$ as
\begin{equation}
u_{\bar{i}i}(\xbf,t) = \sum_{n \geq 1} \sum_{P \in \OWcal_n} \ucal_{\bar{i}i P} \ue^{\ui (\kbf_P \cdot \xbf + \omega_P t)}, 
\end{equation}
where the double copy Berends-Giele currents $\ucal_{\bar{i}i P}$ are determined from the recursion relations
\begin{equation}\label{eq:9.5}
\ucal_{\bar{i}i P} = \frac{ \kappa}{s_P} \sum_{P = Q \cup R}\tfrac{1}{2}\big(\ucal_{\bar{j}j Q} k^{\bar{j}}_{R} k^{j}_{R} \ucal_{\bar{i}i R} - k^{\bar{j}}_{R} \ucal_{\bar{i}j R} k^{j}_{Q} \ucal_{\bar{j}i Q}\big).
\end{equation}
Thus the problem of determining the perturbiner expansion has once again been reduced to that of determining the $L_{\infty}$-quasi-isomorphism from $H^{\sbullet}(\Lfrak')$ to $\Lfrak'$. In the next section we shall see how the double copy prescription can be articulated in terms of the perturbiner coefficients \eqref{eq:9.5}. 


\section{Double copy relations for Berends-Giele currents}\label{sec:10}
So far, we have seen that the double copy structure of the non-abelian Navier-Stokes equation is implied by the factorisation of the strict $L_{\infty}$-algebra $\Lfrak$. Our object is now to uncover this structure at the level of perturbiner expansions. What we shall see is that, from the double copy Berends-Giele currents \eqref{eq:9.5}, we can extract numerators that can be written as the ``square'' of the kinematic numerators obtained from the colour-dressed Berends-Giele currents. This result will be exceedingly useful in permitting us to derive a Kawai-Lewellen-Tye type relation, giving the double copy Berends-Giele currents as a sum of products of two colour-stripped Berends-Giele currents.

First of all it is to be remarked that, in close parallel with development described in Section~\ref{sec:6}, the single index double copy Berends-Giele current $\ucal_{\bar{i}i p}$ may be decomposed into its kinematic degrees of freedom according to
\begin{equation}\label{eq:10.1}
\ucal_{\bar{i}i p} = \bar{\varepsilon}_{\bar{i}p} \varepsilon_{i p}. 
\end{equation}
Here it must be recalled that $\bar{\varepsilon}_{\bar{i}p}$ and $\varepsilon_{i p}$ are regarded as the components of two covectors $\bar{\varepsilon}_{p}$ and $\varepsilon_{p}$ in $\RR^3$. Thus it is possible to consider not one but two infinite-dimensional Lie algebras $\bar{\gfrak}'$ and $\gfrak'$ generated respectively by the $\bar{\varepsilon}_{p}$ and the $\varepsilon_{p}$ and with the same Lie bracket as the one defined in  Eq.~\eqref{eq:7.2}. 

The next step is to write down explicitly the Berends-Giele currents up to multiplicity four and compare them with the expressions given in Section~\ref{sec:7}. We set $P = 12$ in Eq.~\eqref{eq:9.5} first. A simple calculation, using Eq.~\eqref{eq:10.1}, gives
\begin{equation}\label{eq:10.2}
\ucal_{\bar{i} i 12} = \frac{\kappa}{2} \Bigg( \frac{ \bar{\varepsilon}_{\bar{i} [1,2]} \varepsilon_{i [1,2]}}{s_{12}}\Bigg),
\end{equation}
where, as before, we have used the notation $\bar{\varepsilon}_{\bar{i} [1,2]} = [\bar{\varepsilon}_{1}, \bar{\varepsilon}_{2}]_{\bar{i}}$ and $\varepsilon_{i [1,2]} = [\varepsilon_{1},\varepsilon_{2}]_{i}$. We can compare this result directly with Eq.~\eqref{eq:7.5}. The equations are identical if we just identify $\bar{\varepsilon}_{\bar{i} [1,2]}$ with $c^{a}_{[1,2]}$, while sending $\lambda$ to $\frac{\kappa}{2}$. Next we take $P =123$ in Eq.~\eqref{eq:9.5}. In this case, the computation is slightly more complicated, but still straightforward. One finds that
\begin{align}\label{eq:10.3}
\ucal_{\bar{i} i 123} =\left(\frac{\kappa}{2}\right)^2  \Bigg( \frac{\bar{\varepsilon}_{\bar{i}[[1,2],3]} \varepsilon_{i [[1,2],3]}}{ s_{12} s_{123}} + \frac{\bar{\varepsilon}_{\bar{i}[[1,3],2]} \varepsilon_{i [[1,3],2]}}{s_{13} s_{123}} + \frac{\bar{\varepsilon}_{\bar{i}[[2,3],1]} \varepsilon_{i [[2,3],1]}}{s_{23} s_{123}}  \Bigg),
\end{align}
where this time we have used $\bar{\varepsilon}_{\bar{i} [[1,2],3]} = [[\bar{\varepsilon}_{1}, \bar{\varepsilon}_{2}],\bar{\varepsilon}_{3}]_{\bar{i}}$, $\bar{\varepsilon}_{\bar{i} [[1,3],2]} = [[\bar{\varepsilon}_{1}, \bar{\varepsilon}_{3}],\bar{\varepsilon}_{2}]_{\bar{i}}$, $\bar{\varepsilon}_{\bar{i} [[2,3],1]} = [[\bar{\varepsilon}_{2}, \bar{\varepsilon}_{3}],\bar{\varepsilon}_{1}]_{\bar{i}}$, and likewise for the unbarred factors. Thus once again we see that this equation is identical to Eq.~\eqref{eq:7.6} except for the fact that $c^{a}_{[[1,2],3]}$, $c^{a}_{[[1,3],2]}$ and $c^{a}_{[[2,3],1]}$ are replaced by $\bar{\varepsilon}_{\bar{i} [[1,2],3]}$, $\bar{\varepsilon}_{\bar{i} [[1,3],2]}$ and $\bar{\varepsilon}_{\bar{i} [[2,3],1]}$, respectively, and $\lambda$ is sent to $\frac{\kappa}{2}$. Finally, we take $P = 1234$ in Eq.~\eqref{eq:9.5}. By following the same arguments used to derive Eq.~\eqref{eq:10.3}, it is not too hard, but somewhat tedious, to show that
\begin{align}\label{eq:10.4}
\begin{split}
\ucal_{\bar{i} i 1234} =\left(\frac{\kappa}{2}\right)^3 \Bigg( & \frac{\bar{\varepsilon}_{\bar{i}[[[1,2],3],4]} \varepsilon_{i [[[1,2],3],4]}}{s_{12} s_{123} s_{1234}} + \frac{\bar{\varepsilon}_{\bar{i}[[[1,2],4],3]} \varepsilon_{i [[[1,2],4],3]}}{s_{12} s_{124} s_{1234}} +\frac{\bar{\varepsilon}_{\bar{i}[[[1,3],2],4]} \varepsilon_{i [[[1,3],2],4]}}{s_{13} s_{123} s_{1234}} \\
&+ \frac{\bar{\varepsilon}_{\bar{i}[[[1,3],4],2]} \varepsilon_{i [[[1,3],4],2]}}{s_{13} s_{134} s_{1234}} +\frac{\bar{\varepsilon}_{\bar{i}[[[1,4],2],3]} \varepsilon_{i [[[1,4],2],3]}}{s_{14} s_{124} s_{1234}} + \frac{\bar{\varepsilon}_{\bar{i}[[[1,4],3],2]} \varepsilon_{i [[[1,4],3],2]}}{s_{14} s_{134} s_{1234}} \\
&+\frac{\bar{\varepsilon}_{\bar{i}[[[2,3],1],4]} \varepsilon_{i [[[2,3],1],4]}}{s_{23} s_{123} s_{1234}} + \frac{\bar{\varepsilon}_{\bar{i}[[[2,3],4],1]} \varepsilon_{i [[[2,3],4],1]}}{s_{23} s_{234} s_{1234}} +\frac{\bar{\varepsilon}_{\bar{i}[[[2,4],1],3]} \varepsilon_{i [[[2,4],1],3]}}{s_{24} s_{124} s_{1234}} \\
& +\frac{\bar{\varepsilon}_{\bar{i}[[[2,4],3],1]} \varepsilon_{i [[[2,4],3],1]}}{s_{24} s_{234} s_{1234}} +\frac{\bar{\varepsilon}_{\bar{i}[[[3,4],1],2]} \varepsilon_{i [[[3,4],1],2]}}{s_{34} s_{134} s_{1234}} +\frac{\bar{\varepsilon}_{\bar{i}[[[3,4],2],1]} \varepsilon_{i [[[3,4],2],1]}}{s_{34} s_{234} s_{1234}} \\
&+\frac{\bar{\varepsilon}_{\bar{i}[[1,2],[3,4]]} \varepsilon_{i [[1,2],[3,4]]}}{s_{12} s_{34} s_{1234}} + \frac{\bar{\varepsilon}_{\bar{i}[[1,3],[2,4]]} \varepsilon_{i [[1,3],[2,4]]}}{s_{13} s_{24} s_{1234}} +\frac{\bar{\varepsilon}_{\bar{i}[[1,4],[2,3]]} \varepsilon_{i [[1,4],[2,3]]}}{s_{14} s_{23} s_{1234}} \Bigg),
\end{split}
\end{align}
where, of course, we have defined $\bar{\varepsilon}_{\bar{i}[[[1,2],3],4]} = [[[\bar{\varepsilon}_1,\bar{\varepsilon}_2],\bar{\varepsilon}_3],\bar{\varepsilon}_4]_{\bar{i}}$, $\bar{\varepsilon}_{\bar{i}[[1,2],[3,4]]} = [[\bar{\varepsilon}_1,\bar{\varepsilon}_2],[\bar{\varepsilon}_3,\bar{\varepsilon}_4]]_{\bar{i}}$, etc., and similarly for the unbarred factors. Upon comparing the expression in Eq.~\eqref{eq:7.7} with that of Eq.~\eqref{eq:10.4}, we see that they are identical except that the colour factors in the numerators are replaced by barred factors, and $\lambda$ is sent to $\frac{\kappa}{2}$. 

Making use of the notation introduced in Section~\ref{sec:7}, we can proceed directly to find a general expression for $\ucal_{\bar{i}i P}$. Either by explicit calculation based on Eq.~\eqref{eq:9.5} or by inference from Eqs.~\eqref{eq:10.2}, \eqref{eq:10.3}  and \eqref{eq:10.4}, we arrive at
\begin{equation}\label{eq:10.5}
\ucal_{\bar{i}i P} = \left( \frac{\kappa}{2}\right)^{\lvert P \rvert -1} \llbracket \bar{\varepsilon}_{\bar{i}} \otimes \varepsilon_{i} \rrbracket \circ b_{\cd}(P).
\end{equation}
We can compare this with the expression we obtained for the colour-dressed Berends-Giele current given by Eq.~\eqref{eq:7.12}. The similarity of these expressions leads us to conclude that they are identical if $c^{a}$ is replaced by $\bar{\varepsilon}_{\bar{i}}$, and if $\lambda$ is sent to $\frac{\kappa}{2}$. This is what we mean when we say that the numerators of the double copy Berends-Giele current  $\ucal_{\bar{i}i P}$ can be built as the ``square'' of the kinematic numerators of the colour-dressed Berends-Giele current $\ucal^{a}_{i P}$. We must point out, however, that from the perspective we have taken here, this is a reflection of the homotopy algebraic treatment which yields the double copy strict $L_{\infty}$-algebra $\Lfrak'$. 


\section{The zeroth copy of the non-abelian Navier-Stokes equation}\label{sec:11}
Having pinned down the double copy of the non-abelian Navier-Stokes equation, we would now like to address its zeroth copy. We proceed in essentially the same way as in the double copy case and first formally define the strict $L_{\infty}$-algebra that governs the dynamics of the theory. 

For starters, unlike the initial formulation of the non-abelian Navier-Stokes equation, to specify the zeroth copy, we need not one but two compact Lie groups $G$ and $\bar{G}$. The corresponding Lie algebras are written $\gfrak$ and $\bar{\gfrak}$. We pick generators $T_{a}$ and $\bar{T}_{\bar{a}}$ for $\gfrak$ and $\bar{\gfrak}$ respectively, and let the associated structure constants be given by $f_{ab}^{\phantom{ab}c}$ and $\bar{f}_{\bar{a}\bar{b}}^{\phantom{\bar{a}\bar{b}}\bar{c}}$. We also let $\Omega^{0}_{\td}(\RR^{3}, \gfrak \otimes \bar{\gfrak})$ be the space of time-dependent $0$-forms on $\RR^3$ with values in the bi-adjoint representation of $G \times \bar{G}$ on $\gfrak \otimes \bar{\gfrak}$. Explicitly, an element $u \in \Omega^{0}_{\td}(\RR^{3}, \gfrak \otimes \bar{\gfrak})$ can be written as $u = u^{a \bar{a}} T_{a} \otimes \bar{T}_{\bar{a}}$ with $u^{a \bar{a}} \in \Omega^{0}_{\td}(\RR^3)$. Hence, we can allow the operator $\frac{\partial}{\partial t} + \nu \delta \ud$ to act on $\Omega^{0}_{\td}(\RR^{3}, \gfrak \otimes \bar{\gfrak})$ by
\begin{equation}
\left( \frac{\partial}{\partial t} + \nu \delta \ud\right) u = \left( \frac{\partial}{\partial t} + \nu \delta \ud\right)u^{a \bar{a}} T_{a} \otimes \bar{T}_{\bar{a}}.
\end{equation}
There is, moreover, a binary operation on $\Omega^{0}_{\td}(\RR^{3}, \gfrak \otimes \bar{\gfrak})$ given by the rule
\begin{equation}\label{eq:11.2}
\lbbar u, v\rbbar =  \gamma f_{bc}^{\phantom{bc}a} \bar{f}_{\bar{b}\bar{c}}^{\phantom{\bar{b}\bar{c}}\bar{a}} u^{b \bar{b}}v^{c \bar{c}} T_{a} \otimes \bar{T}_{\bar{a}},
\end{equation}
where $\gamma$ is a coupling constant. With this understanding, we choose the cochain complex underlying the zeroth copy strict $L_{\infty}$-algebra $\Lfrak''$ to be
$$
\Omega^1_{\td}(\RR^3,\gfrak \otimes \bar{\gfrak} )[-1] \xrightarrow{\frac{\partial}{\partial t}+ \nu \delta \ud} \Omega^1_{\td}(\RR^3 ,\gfrak \otimes \bar{\gfrak} )[-2].
$$
As usual, we reserve $l_1$ to designate the differential. Regarding the bracket $l_2 \colon \Omega^1_{\td}(\RR^3,\gfrak \otimes \bar{\gfrak} )[-1]^{\otimes 2} \to \Omega^1_{\td}(\RR^3 ,\gfrak \otimes \bar{\gfrak} )[-2]$, we simply set
\begin{equation}\label{eq:11.3}
l_2(u,v) = \lbbar u,v \rbbar.
\end{equation}
Since this is evidently skew-symmetric and the graded Jacobi identity is trivially satisfied, the graded vector space $\Lfrak'' = \Omega^1_{\td}(\RR^3,\gfrak \otimes \bar{\gfrak} )[-1] \otimes \Omega^1_{\td}(\RR^3 ,\gfrak \otimes \bar{\gfrak} )[-2]$ is, in effect, a strict $L_{\infty}$-algebra. 

It is now quite straightforward to derive the field equation that governs the dynamics of the zeroth copy. We just have to write down the Maurer-Cartan equation associated to $\Lfrak''$. Using the definitions of the differential $l_1$ and the bracket $l_2$, this equation takes the form 
\begin{equation}\label{eq:11.4}
\frac{\partial u}{\partial t} + \nu \delta \ud u + \tfrac{1}{2}  \lbbar u, u \rbbar = 0. 
\end{equation}
It is crucial to note that, once again, the notation has been chosen to ensure that this equation looks exactly the same as Eqs.~\eqref{eq:2.12b} and \eqref{eq:8.5}. In terms of components, Eq.~\eqref{eq:11.4} is written as
\begin{equation}\label{eq:11.5}
\frac{\partial u^{a \bar{a}}}{\partial t} - \nu \Delta u^{a \bar{a}} + \frac{\gamma}{2} f_{bc}^{\phantom{bc}a} \bar{f}_{\bar{b}\bar{c}}^{\phantom{\bar{b}\bar{c}}\bar{a}} u^{b \bar{b}} u^{c \bar{c}} = 0. 
\end{equation}
This can be regarded as a fluid analog of the bi-adjoint scalar theory (see, for instance, Refs.~\cite{Bjerrum-Bohr:2012kaa,Cachazo:2013gna,Cachazo:2013hca,Cachazo:2013iea}). For this reason it is natural call Eq.~\eqref{eq:11.5} the bi-adjoint Navier-Stokes equation.

We now indicate briefly how to recast the definition of the strict $L_{\infty}$-algebra $\Lfrak''$ in a form which enables us to handle perturbiner expansions. Analogously to what we did in Section~\ref{sec:3}, let us fix infinite multisets of colour indices $(a_p)_{p \geq 1}$ and $(\bar{a}_p)_{p \geq 1}$ associated with the Lie algebras $\gfrak$ and $\bar{\gfrak}$, respectively, as well as an infinite set $(\kbf_p,\omega_p)_{p \geq 1}$ of pairs with $\kbf_p \in \RR^3$ and $\omega_p \in \RR$ and such that $\ui \omega_p + \nu \kbf_p^2 = 0$ for each $p \geq 1$. Denote by $\Ecal^0_{\td}(\RR^3, \gfrak \otimes \bar{\gfrak})$ the space of formal series of the form
\begin{equation}\label{eq:11.6}
u(\xbf,t) = \sum_{n \geq 1} \sum_{P,Q \in \Wcal_n} \ucal_{P \vert Q} \ue^{\ui (\kbf_P \cdot \xbf + \omega_P t)} T_{a_P} \otimes \bar{T}_{\bar{a}_Q},
\end{equation}
where the coefficients $\ucal_{P \vert Q}$ are supposed to vanish unless the word $P$ is a permutation of the word $Q$. We keep on calling the elements of $\Ecal^0_{\td}(\RR^3, \gfrak \otimes \bar{\gfrak})$ colour-stripped perturbiner ansatzs. The next step is to extend the operator $\frac{\partial}{\partial t} + \nu \delta \ud$ and the binary operation $\lbbar , \rbbar$ to $\Ecal^0_{\td}(\RR^3, \gfrak \otimes \bar{\gfrak})$. For this, we use the colour-dressed version of the pertubiner ansatz. We therefore introduce, for each ordered sequence of positive integers $p_1 < p_2 <  \cdots < p_n$, the notations
\begin{align}
\begin{split}
f^{a}_{p_1 p_2 \cdots p_n} &= f_{a_{p_1} a_{p_2}}^{\phantom{a_{p_1} a_{p_2}}}{}^{b}  f_{b a_{p_3}}^{\phantom{b a_{p_3}}}{}^{c} \cdots  f_{d a_{p_{n-1}}}^{\phantom{d a_{p_{n-1}}}}{}^{e}  f_{e a_{p_{n}}}^{\phantom{e a_{p_{n}}}}{}^{a}, \\
\bar{f}^{\bar{a}}_{p_1 p_2 \cdots p_n} &= \bar{f}_{\bar{a}_{p_1} \bar{a}_{p_2}}^{\phantom{\bar{a}_{p_1} \bar{a}_{p_2}}}{}^{\bar{b}}  \bar{f}_{\bar{b} \bar{a}_{p_3}}^{\phantom{\bar{b} \bar{a}_{p_3}}}{}^{\bar{c}} \cdots  \bar{f}_{\bar{d} \bar{a}_{p_{n-1}}}^{\phantom{\bar{d} \bar{a}_{p_{n-1}}}}{}^{\bar{e}} \bar{f}_{\bar{e} \bar{a}_{p_{n}}}^{\phantom{\bar{e} \bar{a}_{p_{n}}}}{}^{\bar{a}},
\end{split}
\end{align}
and define
\begin{equation}\label{eq:11.8}
\ucal_{p_1 p_2 \cdots p_n}^{a \bar{a}} = \sum_{\sigma,\tau} f^{a}_{p_1 p_{\sigma(2)} \cdots p_{\sigma(n)}} \bar{f}^{\bar{a}}_{p_1 p_{\tau(2)} \cdots p_{\tau(n)}} \ucal_{p_1 p_{\sigma(2)} \cdots p_{\sigma(n)} \vert p_1 p_{\tau(2)} \cdots p_{\tau(n)}},
\end{equation}
where the sums extends over all permutations of the set $(2,\dots,n)$. Using the latter, we can rewrite Eq.~\eqref{eq:11.6} as $u(\xbf,t) = u^{a  \bar{a}}(\xbf,t) T_{a} \otimes \bar{T}_{\bar{a}}$, where the coefficients $u^{a  \bar{a}}(\xbf,t)$ are formal series of the form
\begin{equation}
u^{a  \bar{a}}(\xbf,t) = \sum_{n \geq 1} \sum_{P \in \OWcal_n}  \ucal^{a \bar{a}}_P \ue^{\ui (k_P \cdot \xbf + \omega_P t)}. 
\end{equation}
With this expression in hand, it is now straightforward to extend the definition of $\frac{\partial}{\partial t} + \nu \delta \ud$ and $\lbbar , \rbbar$ to all of $\Ecal^0_{\td}(\RR^3, \gfrak \otimes \bar{\gfrak})$. 

In light of the above discussion, the cochain complex underlying the strict $L_{\infty}$-algebra $\Lfrak''$ that encapsulates perturbiner expansions for the bi-adjoint Navier-Stokes equation is
\begin{equation}\label{eq:11.10}
\Ecal^0_{\td}(\RR^3 ,\gfrak \otimes \bar{\gfrak})[-1] \xrightarrow{\frac{\partial}{\partial t}+ \nu \delta \ud} \Ecal^0_{\td}(\RR^3  ,\gfrak \otimes \bar{\gfrak})[-2].
\end{equation}
That is to say, the differential $l_1$ is the operator $\frac{\partial}{\partial t} + \nu \delta \ud$ acting on $\Ecal^0_{\td}(\RR^3, \gfrak \otimes \bar{\gfrak})$. As for the bracket $l_2 \colon \Ecal^0_{\td}(\RR^3 ,\gfrak \otimes \bar{\gfrak})[-1]^{\otimes 2} \to \Ecal^0_{\td}(\RR^3  ,\gfrak \otimes \bar{\gfrak})[-2]$ it is again determined by the binary operation $\lbbar , \rbbar$. 

Before leaving this section it will be well to comment on the homotopy algebraic structure implicit in the zeroth copy prescription. We assert that $\Lfrak''$ can be obtained by replacing the kinematic factor $\Kin$ in the factorisation \eqref{eq:6.15} of $\Lfrak$ with the colour factor $\bar{\gfrak}$, while sending $\lambda$ to $\gamma$. In other words,
\begin{equation}
\Lfrak'' = \gfrak \otimes (\bar{\gfrak} \otimes \Scal).  
\end{equation}
To justify this assertion, we first observe that $\Ecal^0_{\td}(\RR^3, \gfrak \otimes \bar{\gfrak})$ can be identified with $\gfrak \otimes \bar{\gfrak}  \otimes \Ecal^0_{\td}(\RR^3)$ and hence we may write each element $u(\xbf,t) \in \Ecal^0_{\td}(\RR^3, \gfrak \otimes \bar{\gfrak})$ in the form $u(\xbf,t) = T_{a} \otimes \bar{T}_{\bar{a}} \otimes u^{a \bar{a}}(\xbf,t)$ with $u^{a \bar{a}}(\xbf,t) \in \Ecal^{0}_{\td}(\RR^3)$. With this identification, it follows at once that $\Lfrak'' = \gfrak \otimes (\bar{\gfrak} \otimes \Scal)$ as graded vector spaces.  Furthermore, calling to mind the definition of the differential $\mu_1$ on $\Scal$, we have
$$
l_1 (T_{a} \otimes \bar{T}_{\bar{a}} \otimes u^{a \bar{a}}(\xbf,t)) = T_{a} \otimes \bar{T}_{\bar{a}} \otimes \left( \frac{\partial}{\partial t} + \nu \delta \ud \right) u^{a \bar{a}}(\xbf,t) = T_{a} \otimes \bar{T}_{\bar{a}} \otimes  \mu_1 (u^{a \bar{a}}(\xbf,t)).
$$
So there only remains to verify that
\begin{equation}\label{eq:11.12}
l_2 (T_{a} \otimes \bar{T}_{\bar{a}} \otimes u^{a \bar{a}}(\xbf,t),T_{b} \otimes \bar{T}_{\bar{b}} \otimes v^{b \bar{b}}(\xbf,t)) = [T_{a},T_{b}] \otimes [\bar{T}_{\bar{a}}, \bar{T}_{\bar{b}}] \otimes \mu_2 (u^{a \bar{a}}(\xbf,t),v^{b \bar{b}}(\xbf,t))). 
\end{equation}
To begin with, from Eqs.~\eqref{eq:11.2}  and \eqref{eq:11.3} we see that
\begin{align}\label{eq:11.13}
\begin{split}
l_2 (T_{a} \otimes \bar{T}_{\bar{a}} \otimes u^{a \bar{a}}(\xbf,t),T_{b} \otimes \bar{T}_{\bar{b}} \otimes v^{b \bar{b}}(\xbf,t)) &= T_{c} \otimes \bar{T}_{\bar{c}} \otimes \gamma f_{ab}^{\phantom{ab}c} \bar{f}_{\bar{a}\bar{b}}^{\phantom{\bar{a}\bar{b}}\bar{c}} u^{a \bar{a}}(\xbf,t) v^{b \bar{b}}(\xbf,t)  \\
&= [T_{a},T_{b}] \otimes [\bar{T}_{\bar{a}}, \bar{T}_{\bar{b}}] \otimes \gamma u^{a \bar{a}}(\xbf,t) v^{b \bar{b}}(\xbf,t). 
\end{split}
\end{align}
If we denote by $\ucal^{a \bar{a}}_P$ and $\vcal^{b \bar{b}}_Q$ the coefficients in the expansions of $u^{a \bar{a}}(\xbf,t)$ and $v^{b \bar{b}}(\xbf,t)$, respectively, then Eq.~\eqref{eq:11.13} can also be written in the form
\begin{align}\label{eq:11.14}
\begin{split}
&l_2 (T_{a} \otimes \bar{T}_{\bar{a}} \otimes u^{a \bar{a}}(\xbf,t),T_{b} \otimes \bar{T}_{\bar{b}} \otimes v^{b \bar{b}}(\xbf,t)) \\
&\qquad \qquad = [T_{a},T_{b}] \otimes [\bar{T}_{\bar{a}}, \bar{T}_{\bar{b}}] \otimes \sum_{n \geq 1} \sum_{P \in \OWcal_n} \Bigg( \sum_{P = Q \cup R} \gamma \ucal^{a \bar{a}}_Q \vcal^{b \bar{b}}_R \Bigg) \ue^{\ui (\kbf_P \cdot \xbf + \omega_P t)}.
\end{split}
\end{align}
On the other hand, from Eq.~\eqref{eq:6.5}, but with $\lambda$ replaced with $\gamma$, we get
\begin{equation}\label{eq:11.15}
 \mu_2 (u^{a \bar{a}}(\xbf,t),v^{b \bar{b}}(\xbf,t))) = \sum_{n \geq 1} \sum_{P \in \OWcal_n} \Bigg( \sum_{P = Q \cup R} \gamma \ucal^{a \bar{a}}_Q \vcal^{b \bar{b}}_R \Bigg) \ue^{\ui (\kbf_P \cdot \xbf + \omega_P t)}. 
\end{equation}
Combining Eq.~\eqref{eq:11.14} and Eq.~\eqref{eq:11.15}, we arrive at Eq.~\eqref{eq:11.12}, as we wished to check. 


\section{Multiparticle solution to the bi-adjoint Navier-Stokes equation}\label{sec:12}
We now briefly address the problem of finding a multiparticle solution to the bi-adjoint Navier-Stokes equation. Just as we have done above, we can build up such solution in the form of a perturbiner expansion and comes in two flavours:~a colour-stripped version and a colour-dressed version.  

We begin as usual by considering the minimal $L_{\infty}$-structure on the cohomology $H^{\sbullet}(\Lfrak'')$ of the zeroth copy strict $L_{\infty}$-algebra $\Lfrak''$. For this, we need to specify a projection $p \colon \Lfrak'' \to H^{\sbullet}(\Lfrak'')$ and a contracting homotopy $h \colon \Lfrak'' \to \Lfrak''$. In the first place, as a consequence of the definition \eqref{eq:11.10}, the cochain complex underlying $H^{\sbullet}(\Lfrak'')$ is 
$$
\ker (l_1)[-1] \xrightarrow{\phantom{aa} 0 \phantom{aa}} \coker (l_1)[-2],
$$
which is formally identical with those found in Sections~\ref{sec:4} and \ref{sec:9}. We then proceed as before, extending the Wyld operator $G^{\uW}$ to a linear operator $G^{\uW} \colon \Ecal^0_{\td}(\RR^3, \gfrak \otimes \bar{\gfrak}) \to \Ecal^0_{\td}(\RR^3, \gfrak \otimes \bar{\gfrak})$, taking the projection $p^{(1)} \colon \Ecal^0_{\td}(\RR^3, \gfrak \otimes \bar{\gfrak}) \to \ker(l_1)$ to be given by
\begin{equation}
p^{(1)} = \id_{\Ecal^0_{\td}(\RR^3, \gfrak \otimes \bar{\gfrak})} - G^{\uW} \circ l_1,
\end{equation}
and the projection $p^{(2)} \colon \Ecal^0_{\td}(\RR^3, \gfrak \otimes \bar{\gfrak}) \to \coker(l_1)$ to be given by the quotient map. Using these, one can deduce that $h^{(2)}= G^{\uW}$ is the only non-zero component of the contracting homotopy $h$. 

We are now in a position to consider the perturbiner expansion for the bi-adjoint Navier-Stokes equation. First we discuss the colour-stripped version. On that account, we select a Maurer-Cartan element $u'(\xbf,t) \in H^1(\Lfrak'')=\ker(l_1)$ of the form
\begin{equation}
u'(\xbf,t) = \sum_{p,q \geq 1} \ucal_{p \vert q} \ue^{\ui (\kbf_P \cdot \xbf + \omega_P t)} T_{a_p} \otimes \bar{T}_{\bar{a}_q}. 
\end{equation}
Then the colour-stripped perturbiner expansion may be defined as the Maurer-Cartan element $u(\xbf,t)$ of $\Lfrak''$ given by the formula
\begin{equation}\label{eq:12.3}
u(\xbf,t) = \sum_{n \geq 1} \frac{1}{n!} f_n (u'(\xbf,t), \dots, u'(\xbf,t)). 
\end{equation}
Using precisely the same technique as we used in Section~\ref{sec:4}, we can evaluate the right hand side of Eq.~\eqref{eq:12.3}. We obtain
\begin{equation}
u(\xbf,t) = \sum_{n \geq 1} \sum_{P,Q \in \Wcal_n}  \ucal_{P \vert Q} \ue^{\ui(\kbf_P \cdot \xbf + \omega_P t)} T_{a_P} \otimes \bar{T}_{\bar{a}_P},
\end{equation}
where the coefficients $\ucal_{P \vert Q}$ are determined from the recursion relations
\begin{equation}\label{eq:12.5}
\ucal_{P \vert Q} = \frac{\gamma}{s_P} \sum_{P = RS} \sum_{Q= TU} (\ucal_{R \vert T} \ucal_{S \vert U} - \ucal_{S \vert T} \ucal_{R \vert U}).
\end{equation}
Notice that, because of the antisymmetry of the right-hand side of Eq.~\eqref{eq:12.5} under interchange of the words $R$ and $S$ and $T$ and $U$, these coefficients obey the shuffle constraint $\ucal_{P \shuffle Q \vert R} = 0$. In view of this and the similarity already remarked with the bi-adjoint scalar theory, we adopt the terminology of Ref.~\cite{Mafra:2016ltu}, and refer to the $\ucal_{P \vert Q}$ as the Berends-Giele double currents. It is also worth pointing out that, if we choose $\ucal_{p \vert q} = \delta_{pq}$, the recursion relation in Eq.~\eqref{eq:12.5} can be rewritten in terms of the colour-stripped Berends-Giele map $b_{\cs}$ as
\begin{equation}\label{eq:12.6}
\ucal_{P \vert Q} = \gamma ( P, b_{\cs}(Q)), 
\end{equation}
where $( , )$ stands for the scalar product of words. This formula will play a crucial role in our further treatment. 

We next discuss the colour-dressed version of the perturbiner expansion. To that effect, we pick a Maurer-Cartan element $u'(\xbf,t) \in H^{1}(\Lfrak'') = \ker(l_1)$ with components of the form
\begin{equation}
u'{}^{a \bar{a}}(\xbf,t) = \sum_{p \geq 1} \ucal^{a \bar{a}}_{p} \ue^{\ui (\kbf_p \cdot \xbf  + \omega_p t)}.  
\end{equation}
One may then define the colour-dressed perturbiner expansion $u(\xbf, t) \in \Lfrak''$ by exactly the same formula as Eq.~\eqref{eq:12.3}. The problem is to work out its components. By a virtually identical calculation to that followed in Section~\ref{sec:5}, we find that
\begin{equation}
u^{a \bar{a}}(\xbf,t) = \sum_{n \geq 1} \sum_{P \in \OWcal_n}  \ucal^{a \bar{a}}_P \ue^{\ui (\kbf_P \cdot \xbf + \omega_P t)}, 
\end{equation}
where the coefficients $\ucal^{a \bar{a}}_P$, which we shall refer to as the zeroth copy Berends-Giele currents, are determined from the recursion relations
\begin{equation}\label{eq:12.9}
\ucal^{a \bar{a}}_P = \frac{\gamma}{s_P} \sum_{P = Q \cup R} \tfrac{1}{4} \tilde{f}_{bc}^{\phantom{bc}a}  \tilde{\bar{f}}_{\bar{b}\bar{c}}^{\phantom{\bar{b}\bar{c}}\bar{a}} \ucal^{b \bar{b}}_Q \ucal^{c \bar{c}}_R.  
\end{equation}
Here we found it convenient, as in Section~\ref{sec:5}, to introduce the notations $\tilde{f}_{bc}^{\phantom{bc}a} = - 2 \ui f_{bc}^{\phantom{bc}a}$ and $\tilde{\bar{f}}_{\bar{b}\bar{c}}^{\phantom{\bar{b}\bar{c}}\bar{a}} = -2 \ui \bar{f}_{\bar{b}\bar{c}}^{\phantom{\bar{b}\bar{c}}\bar{a}}$. Of course, this equation can be cast in a form analogous to Eq.~\eqref{eq:7.12}. To this end, for each bracketed word $\ell[P] = \ell[p_1p_2 \cdots p_n]$ of length $n$, we set
\begin{align}\label{eq:12.10}
\begin{split}
c^{a}_{\ell[P]} &= \tilde{f}_{a_{p_1} a_{p_2}}^{\phantom{a_{p_1} a_{p_2}}}{}^{b}  \tilde{f}_{b a_{p_3}}^{\phantom{b a_{p_3}}}{}^{c} \cdots  \tilde{f}_{d a_{p_{n-1}}}^{\phantom{d a_{p_{n-1}}}}{}^{e}  \tilde{f}_{e a_{p_{n}}}^{\phantom{e a_{p_{n}}}}{}^{a}, \\
\bar{c}^{\bar{a}}_{\ell[P]} &= \tilde{\bar{f}}_{\bar{a}_{p_1} \bar{a}_{p_2}}^{\phantom{\bar{a}_{p_1} \bar{a}_{p_2}}}{}^{\bar{b}}  \tilde{\bar{f}}_{\bar{b} \bar{a}_{p_3}}^{\phantom{\bar{b} \bar{a}_{p_3}}}{}^{\bar{c}} \cdots  \tilde{\bar{f}}_{\bar{d} \bar{a}_{p_{n-1}}}^{\phantom{\bar{d} \bar{a}_{p_{n-1}}}}{}^{\bar{e}} \tilde{\bar{f}}_{\bar{e} \bar{a}_{p_{n}}}^{\phantom{\bar{e} \bar{a}_{p_{n}}}}{}^{\bar{a}},
\end{split}
\end{align}
with the conventions $c^{a}_{p} = \delta^{a}_{\phantom{a} a_p}$ and $\bar{c}^{\bar{a}}_{p} = \delta^{\bar{a}}_{\phantom{\bar{a}} \bar{a}_p}$. We further define $c^{a}_{[\ell[P],\ell[Q]]} = \tilde{f}_{bc}^{\phantom{bc}a} c^{b}_{\ell[P]} c^{c}_{\ell[Q]}$ and $\bar{c}^{\bar{a}}_{[\ell[P],\ell[Q]]} = \tilde{\bar{f}}_{\bar{b}\bar{c}}^{\phantom{\bar{b}\bar{c}}\bar{a}} \bar{c}^{\bar{b}}_{\ell[P]} \bar{c}^{\bar{c}}_{\ell[Q]}$ for every pair of bracketed words $\ell[P]$ and $\ell[Q]$. Then it is an easy matter to show that, in the notation of Section~\ref{sec:7}, the recursion relation of Eq.~\eqref{eq:12.9} can be rewritten more simply as
\begin{equation}\label{eq:12.11}
\ucal^{a \bar{a}}_P = \left( \frac{\gamma}{4}\right)^{\lvert P\rvert - 1} \llbracket c^{a} \otimes \bar{c}^{\bar{a}}\rrbracket \circ b_{\cd}(P). 
\end{equation}
By comparing this with Eq.~\eqref{eq:7.12}, we see that the effect of the zeroth copy construction at the level of colour-dressed perturbiner expansions is equivalent to replace $\varepsilon_i$ by $\bar{c}^{\bar{a}}$ and send $\lambda$ to $\frac{\gamma}{4}$. Once again this is a reflection of the homotopy algebraic perspective we have adopted to produce the zeroth copy strict $L_{\infty}$-algebra $\Lfrak''$.\footnote{It is important to notice that the realisation of the $L_{\infty}$-algebra $\Lfrak''$ as a zeroth copy required $\lambda$ to be sent to $\gamma$ and not to~$\frac{\gamma}{4}$. This is because in the argument presented in Section~\ref{sec:11} we worked with the structure constants $f_{ab}^{\phantom{ab}c}$ and $\bar{f}_{\bar{a}\bar{b}}^{\phantom{\bar{a}\bar{b}}\bar{c}}$ instead of the rescaled ones $\tilde{f}_{ab}^{\phantom{ab}c}$ and $\tilde{\bar{f}}_{\bar{a}\bar{b}}^{\phantom{\bar{a}\bar{b}}\bar{c}}$.}


\section{Scattering amplitudes for the non-abelian and tensor Navier-Stokes equations}\label{sec:13}
Tree-level scattering amplitudes of ``fluid quanta'' described by the non-abelian Navier-Stokes equation have been explored   in Ref.~\cite{Cheung:2020djz}. In this section we use the pioneering results of Berends and Giele in Ref.~\cite{Berends:1987me} to construct multiparticle generalisations of such amplitudes. Naturally this approach also allows for a multiparticle description of tree-level scattering amplitudes of ``bi-fluid quanta'' described by the tensor Navier-Stokes equation. We shall learn in the next section that, through the double copy, the latter can be expressed as sums of products of the former.

Following the path of the previous sections, we shall start by examining the colour-ordered partial amplitudes associated to the non-abelian Navier-Stokes equation which are directly related to the colour-stripped Berends-Giele currents. Throughout the discussion, we let $P$ denote an arbitrary but fixed permutation of $2 3 \cdots (n-1)$. The external data for a scattering amplitude involve a specification of a momentum $\kbf_p$ and an energy $\omega_p$ for each of the $n$ fluid quanta, subject to the dispersion relation $\ui \omega_p + \nu \kbf_p^2 = 0$ and momentum conservation $-\kbf_n = \sum_{p=1}^{n-1} \kbf_p$. One often uses the phrase ``on-shell constraint'' to refer to the first of these two conditions. We also recall that the single index colour-stripped Berends-Giele currents $\ucal_{ip} = \varepsilon_{i p}$ satisfy the transversality condition $\varepsilon_{p}^{ \sharp} \cdot \kbf_p = 0$, so that they may be assimilated to polarisations of fluid quanta. With these general considerations in mind and motivated by the Berends-Giele prescription, the tree-level colour-ordered partial amplitude for the scattering of the $n$ fluid quanta is defined through the formula
\begin{equation}\label{eq:13.1}
A^{\mathrm{tree}} (1Pn) = s_{1P} \ucal_{i 1P} \varepsilon^{i}_{n}. 
\end{equation}
Here the factor $s_{1P}$ is inserted to cancel the $1/s_{1P}$ pole inside $\ucal_{i 1P}$. It is also worth stressing that Eqs.~\eqref{eq:4.14} and \eqref{eq:4.15} imply that the right-hand side of Eq.~\eqref{eq:13.1} is manifestly energy independent in the sense that it depends only on dot products of momenta and polarisations. Finally, we note that, as a consequence of the shuffle constraint for the colour-stripped Berends-Giele currents, the partial amplitude given by Eq.~\eqref{eq:13.1} manifestly satisfies the Kleiss-Kuijf relations
\begin{equation}
A^{\mathrm{tree}} (Q1Rn) = (-1)^{\lvert Q \rvert} A^{\mathrm{tree}} (1 (\bar{Q} \shuffle R) n),
\end{equation}
where the words $Q$ and $R$ involve the labels $2,\dots, n-1$. As usual, these relations allow for a basis of $(n-2)!$ tree-level colour-ordered partial amplitudes with letters $1$ and $n$ held fixed (see, for instance, Refs.~\cite{DelDuca:1999rs} and \cite{Bern:2019prr}). 

Our object now is to consider the amplitudes associated to the non-abelian Navier-Stokes equation that are linked to the colour-dressed Berends-Giele currents. The formula to evaluate such amplitudes turns out to be a straightforward generalisation of the one considered in the colour-stripped case. Indeed, following the analogy developed in Ref.~\cite{Mizera:2018jbh}, the full amplitude for the scattering of $n$ fluid quanta can be determined using
\begin{equation}
\Acal^{\mathrm{tree}}_n = s_{12\cdots (n-1)} \ucal_{i 12 \cdots (n-1)}^{a} \ucal^{i}_{a n} 
\end{equation}
or, equivalently, in view of Eq.~\eqref{eq:7.1},
\begin{equation}\label{eq:13.4}
\Acal^{\mathrm{tree}}_n = s_{12\cdots (n-1)} \ucal_{i 12 \cdots (n-1)}^{a_n} \varepsilon^{i}_{n}. 
\end{equation}
We may also express this another way. We first note that by use of Eq.~\eqref{eq:7.3} we can rewrite Eq.~\eqref{eq:3.5} in the form
\begin{equation}\label{eq:13.5}
\ucal^{a}_{i 12 \cdots (n-1)} = \left( \tfrac{\ui}{2}\right)^{n-2} \sum_{P} c^{a}_{\ell[1P]} \ucal_{i 1 P}, 
\end{equation}
where the sum over $P$ represents the sum over all permutations of $2 3 \cdots (n-1)$. Substituting Eq.~\eqref{eq:13.5} back into \eqref{eq:13.4}, we obtain
\begin{equation}
\Acal^{\mathrm{tree}}_n = \left( \tfrac{\ui}{2}\right)^{n-2} \sum_{P} c^{a_n}_{\ell[1P]} s_{12\cdots (n-1)} \ucal_{i 1 P} \varepsilon^{i}_{n} = \left( \tfrac{\ui}{2}\right)^{n-2} \sum_{P} c^{a_n}_{\ell[1P]} s_{1P} \ucal_{i 1 P} \varepsilon^{i}_{n},
\end{equation}
where the last equality follows from the identity $s_{12\cdots (n-1)} = s_{1P}$ for $P$ a permutation of $2 3 \cdots (n-1)$. From the definition \eqref{eq:13.1}, we conclude then that
\begin{equation}\label{eq:13.7}
\Acal^{\mathrm{tree}}_n = \left( \tfrac{\ui}{2}\right)^{n-2} \sum_{P} c^{a_n}_{\ell[1P]} A^{\mathrm{tree}}(1Pn). 
\end{equation}
The significance of this equation is that we can always decompose the tree-level scattering amplitude $\Acal_n$ in terms of colour factors $c^{a_n}_{\ell[1P]}$ and tree-level colour-ordered partial amplitudes $A^{\mathrm{tree}}(1Pn)$, which, as we have seen, contain all kinematical information. 

It remains to say a word about amplitudes associated to the tensor Navier-Stokes equation which are obtainable directly from the double copy Berends-Giele currents. The external data to specify them is as above:~a momentum $\kbf_p$ and an energy $\omega_p$ for each of the $n$ bi-fluid quanta, subject to the on-shell constraint $\ui \omega_p + \nu \kbf_p^2=0$ and momentum conservation $-\kbf_{n}= \sum_{p=1}^{n-1}\kbf_p$. The above description then shows (or rather, suggests) that the amplitude for the scattering of the $n$ bi-fluid quanta can be represented as
\begin{equation}
\Mcal^{\mathrm{tree}}_n = s_{12 \cdots (n-1)} \ucal_{\bar{i}i 12\cdots (n-1)} \ucal^{\bar{i} i}_{n}, 
\end{equation}
or, equivalently, using Eq.~\eqref{eq:10.1},
\begin{equation}\label{eq:13.9}
\Mcal^{\mathrm{tree}}_n = s_{12 \cdots (n-1)} \ucal_{\bar{i}i 12\cdots (n-1)} \bar{\varepsilon}^{\bar{i}}_n \varepsilon^{i}_n. 
\end{equation}
However, for the analogy with scattering of fluid quanta to be complete, we would like the amplitude $\Mcal^{\mathrm{tree}}_n$ to be expressed in terms of a sum of colour-ordered partial amplitudes times kinematic factors. This will be shown to follow from the basic property \eqref{eq:10.5} of double copy Berends-Giele currents.  


\section{Kawai-Lewellen-Tye relations}\label{sec:14}
We must turn now to a more detailed examination of the foregoing proposed form of the scattering amplitude of bi-fluid quanta. 
We shall show that these amplitudes can be expressed either in terms of a sum over products of colour-ordered partial amplitudes and master numerators that depend on the kinematics variables, or in terms of a sum over products of pairs of colour-ordered partial amplitudes ``glued'' together by kinematic factors contained in the so-called momentum kernel. The latter of these two provides an analogue of the Kawai-Lewellen-Tye relations that link products of Yang-Mills amplitudes to gravity amplitudes at tree level \cite{Bern:2019prr,Bern:2010yg,Bjerrum-Bohr:2010diw,Bjerrum-Bohr:2010pnr,Fu:2012uy,Fu:2017uzt,Geyer:2017ela}. To get somewhat cleaner looking expressions, throughout our discussion we shall set $\lambda = 1$, $ \kappa = 1$ and $\gamma = 1$.

We begin the treatment by determining a simple relationship between the colour-dressed Berends-Giele map and the Berends-Giele double currents. The first thing to notice is that, since $b_{\cd}(12\cdots (n-1))$ is a Lie polynomial of length $n-1$, it may be expanded as
\begin{equation}\label{eq:14.1}
b_{\cd}(12\cdots (n-1)) = \sum_{P} \left( 1P, b_{\cd}(12\cdots (n-1))\right) \ell[1P],
\end{equation}
where the sum runs over all permutations of $23 \cdots (n-1)$. Referring back to Eq.~\eqref{eq:12.11}, we see, then, that
\begin{equation}\label{eq:14.2}
\ucal^{a \bar{a}}_{12\cdots (n-1)} = \left(\tfrac{1}{4}\right)^{n-2} \sum_P c^{a}_{\ell[1P]} \left( 1P, b_{\cd}(12\cdots (n-1))\right) \bar{c}^{\bar{a}}_{\ell[1P]}. 
\end{equation}
On the other hand, making use of Eqs.~\eqref{eq:11.8} and \eqref{eq:12.10}, we can easily obtain 
\begin{equation}\label{eq:14.3}
\ucal^{a \bar{a}}_{12\cdots (n-1)}  = (-1)^{n}\left(\tfrac{1}{4}\right)^{n-2} \sum_P c^{a}_{\ell[1P]} \Bigg( \sum_Q  \ucal_{1P \vert 1 Q} \bar{c}^{\bar{a}}_{\ell[1Q]} \Bigg).
\end{equation}
In order that Eqs.~\eqref{eq:14.2} and \eqref{eq:14.3} both hold we must have
\begin{equation}\label{eq:14.4}
\left( 1P, b_{\cd}(12\cdots (n-1))\right) \bar{c}^{\bar{a}}_{\ell[1P]} =(-1)^{n} \sum_Q  \ucal_{1P \vert 1 Q} \bar{c}^{\bar{a}}_{\ell[1Q]}. 
\end{equation}
And for Eq.~\eqref{eq:14.4} to be true, we must have
\begin{equation}\label{eq:14.5}
\left( 1P, b_{\cd}(12\cdots (n-1))\right) \ell[1P] = (-1)^{n} \sum_Q  \ucal_{1P \vert 1 Q} \ell[1Q].
\end{equation}
This is the expression that we want. It shows that the coefficients in the expansion of the right-hand side as a Lie polynomial of length $n-1$ are, modulo the global sign $(-1)^{n}$, given by the Berends-Giele double currents $\ucal_{1P \vert 1 Q}$. 

We next explore some of the consequences of the above relation. To begin with, for any permutation $P$ of $23 \cdots (n-1)$, $b_{\cs}(1P)$ may be expanded analogously to Eq.~\eqref{eq:14.1} as
\begin{equation}
b_{\cs}(1P) = \sum_Q (1Q,b_{\cs}(1P)) \ell[1Q],
\end{equation}
which, on use of Eq.~\eqref{eq:12.6} and the identity $\ucal_{1Q \vert 1P} =  \ucal_{1P \vert 1Q}$, becomes
\begin{equation}
b_{\cs}(1P) = \sum_Q \ucal_{1P \vert 1Q} \ell[1Q]. 
\end{equation}
Plugging this back into Eq.~\eqref{eq:4.15}, one finds that
\begin{equation}\label{eq:14.8}
\ucal_{i 1P} = \sum_Q \ucal_{1P \vert 1Q} \varepsilon_{i \ell[1Q]}. 
\end{equation}
Alternatively, we could of course use Eq.~\eqref{eq:14.5} to express Eq.~\eqref{eq:14.8} as
\begin{equation}\label{eq:14.9}
\ucal_{i 1P} = (-1)^{n} \left( 1P, b_{\cd}(12\cdots (n-1))\right) \varepsilon_{i \ell[1P]}. 
\end{equation}
Now let us go back to Eq.~\eqref{eq:10.5}. Using Eqs.~\eqref{eq:14.1} and \eqref{eq:14.9}, this is
\begin{equation}\label{eq:14.10}
\ucal_{\bar{i} i 12 \cdots (n-1)} = (-1)^{n}\left(\tfrac{1}{2}\right)^{n-2} \sum_P \bar{\varepsilon}_{\bar{i} \ell[1P]} \ucal_{i 1 P}. 
\end{equation}
Inserting this result into Eq.~\eqref{eq:13.9} gives then
\begin{equation}\label{eq:14.11}
\Mcal^{\mathrm{tree}}_n = (-1)^{n}\left(\tfrac{1}{2}\right)^{n-2} \sum_P \bar{\varepsilon}_{\bar{i} \ell[1P]} \bar{\varepsilon}^{\bar{i}}_n s_{12\cdots(n-1)} \ucal_{i 1P} \varepsilon^{i}_n = (-1)^{n}\left(\tfrac{1}{2}\right)^{n-2} \sum_P \bar{\varepsilon}_{\bar{i} \ell[1P]} \bar{\varepsilon}^{\bar{i}}_n s_{1P} \ucal_{i 1P} \varepsilon^{i}_n, 
\end{equation}
where, as before, we have used the identity $s_{12\cdots(n-1)} = s_{1P}$ for $P$ a permutation of $23\cdots (n-1)$. This leads us to introduce the coefficients 
\begin{equation}
\bar{n}_{\ell[1P]n} = \bar{\varepsilon}_{\bar{i} \ell[1P]} \bar{\varepsilon}^{\bar{i}}_n,
\end{equation}
to which we will refer to as master numerators.\footnote{The terminology ``master numerators'' is taken from the literature in colour-kinematics duality. They are associated with ``half-ladder'' diagrams which are characterised by a fixed choice of endpoints $1$ and $n$ as well as permutations of the remaining legs $2,3,\dots,n-1$.}  By substituting these in Eq.~\eqref{eq:14.11}, and remembering the definition \eqref{eq:13.1}, we finally obtain
\begin{equation}
\Mcal^{\mathrm{tree}}_n = (-1)^{n}\left(\tfrac{1}{2}\right)^{n-2} \sum_P \bar{n}_{\ell[1P]n} A^{\mathrm{tree}}(1Pn). 
\end{equation}
A comparison with Eq.~\eqref{eq:13.7} shows a deep underlying similarity between the tree-level scattering amplitudes $\Acal^{\mathrm{tree}}_n$ and $\Mcal^{\mathrm{tree}}_n$. We need merely replace the colour factor  $c^{a_n}_{\ell[1P]}$ by the master numerator $\bar{n}_{\ell[1P]n}$ to obtain the same looking decomposition in terms of the tree-level colour-ordered partial amplitudes $A^{\mathrm{tree}}(1Pn)$. Needless to say, this circumstance is a proper manifestation of the color-kinematics duality that was described in Section~\ref{sec:7}.

We are now in a position to derive the Kawai-Lewellen-Tye relations. In fact, we shall prove a more general Kawai-Lewellen-Tye type relation which expresses double copy Berends-Giele currents as sums of products of colour-stripped Berends-Giele currents. The key ingredient here is the introduction of a matrix of objects $S(P\vert Q)_p$, labelled by permutations $P$ and $Q$ of $23\cdots (n-1)$ and a positive integer $p$, which is known in the literature as the momentum kernel, and which works as an inverse for the matrix of objects whose entries are the Berends-Giele double currents $\ucal_{pP \vert pQ}$. This latter statement translates to the algebraic relations
\begin{equation}\label{eq:14.14}
\sum_{R} \ucal_{pP \vert pR} S(R \vert Q)_p  = \sum_{R} S(P \vert R)_p \ucal_{pR \vert pQ} = \delta_{P,Q}. 
\end{equation}
The existence of such an inverse can be established by adapting the construction of Ref.~\cite{Frost:2020eoa} to the present context. Indeed, it is possible to demonstrate that a recursive formula for its entries is given by
\begin{align}
\begin{split}
S(\varnothing \vert \varnothing)_p &= 1, \\
S(Pq \vert QqR)_p &= 2 \nu \kbf_{q} \cdot \kbf_{pQ} S(P \vert QR)_p. 
\end{split}
\end{align}
With this implement at our disposal we may next proceed with the derivation. Let us write $\bar{\ucal}_{\bar{i} 1P}$ for the colour-stripped Berends-Giele current associated with the kinematic numerator $\bar{\varepsilon}_{\bar{i} \ell[1P]}$. We have, according to Eq.~\eqref{eq:14.8}, that
\begin{equation}
\bar{\ucal}_{\bar{i} 1P} = \sum_{Q} \ucal_{1P \vert 1Q} \bar{\varepsilon}_{\bar{i} \ell[1Q]}. 
\end{equation}
With Eq.~\eqref{eq:14.14}, this tells us that 
\begin{equation}
\bar{\varepsilon}_{\bar{i} \ell[1Q]} = \sum_{P} S(Q \vert P)_1 \bar{\ucal}_{\bar{i} 1P}. 
\end{equation}
Substituting this back into Eq.~\eqref{eq:14.10}, and using the property $S(Q \vert P)_1 = S(P \vert Q)_1$, then gives
\begin{equation}
\ucal_{\bar{i} i 12 \cdots (n-1)} = (-1)^{n}\left(\tfrac{1}{2}\right)^{n-2} \sum_{P,Q} \bar{\ucal}_{\bar{i} 1P} S(P \vert Q)_1 \ucal_{i 1 Q}.
\end{equation}
This is the equation that we desire. It is in a perfect agreement with the proposal put forward in Ref.~\cite{Mizera:2018jbh}. If we now enter all this information into Eq.~\eqref{eq:13.9}, we find that
\begin{equation}
\Mcal^{\mathrm{tree}}_n = (-1)^{n}\left(\tfrac{1}{2}\right)^{n-2} \sum_{P,Q} \bar{\ucal}_{\bar{i} 1P} \bar{\varepsilon}^{\bar{i}}_n S(P \vert Q)_1 s_{1Q} \ucal_{i 1 Q} \varepsilon^{i}_n,
\end{equation}
which, upon use of the definition \eqref{eq:13.1}, yields up
\begin{equation}\label{eq:14.20}
\Mcal^{\mathrm{tree}}_n = (-1)^{n}\left(\tfrac{1}{2}\right)^{n-2} \sum_{P,Q} \frac{1}{s_{1P}} \bar{A}^{\mathrm{tree}}(1Pn) S(P \vert Q)_1 A^{\mathrm{tree}}(1Qn).
\end{equation}
The equality within this expression gives what is known as the $(n-2)!$ version of the Kawai-Lewellen-Tye relations (see Ref.~\cite{Bjerrum-Bohr:2010diw} for further details). More accurately, it can be seen as the natural adaptation of these relations to a nonrelativistic setting.  

This is the result which we have been seeking:~a prescription for realising the double copy directly at amplitude level. Once we know the colour-ordered partial amplitudes for the non-abelian Navier-Stokes equation we can determine automatically the scattering amplitudes for the tensor Navier-Stokes equation via Eq.~\eqref{eq:14.20}.
 

\section{Conclusions}\label{sec:15}
We investigated how the colour-kinematics duality and the double copy of the non-abelian Navier-Stokes equation can be realised in terms of the homotopy algebraic treatment adduced in Ref.~\cite{Borsten:2021hua}. We showed that by using only information that is intrinsic to the factorisation of the strict $L_{\infty}$-algebra encoding the perturbiner expansions for the non-abelian Navier-Stokes equation, one can pull out kinematic numerators from the colour-dressed Berends-Giele currents that obey the same generalised Jacobi identities as the colour factors, thus manifesting a kinematic Lie algebra. Moreover, we showed that the homotopy double copy prescription applied to such factorisation can be matched with the strict $L_{\infty}$-algebra encoding the perturbiner expansions for the tensor Navier-Stokes equation. Armed with this understanding, we went on to explore the implications of the double copy construction at the level of perturbiner coefficients. We showed that the double copy Berends-Giele currents are obtained by replacing the colour factors of the colour-dressed Berends-Giele currents with kinematic factors. In physical terminology, this is what one would refer to as an ``off-shell'' formulation of the double copy. We also learned that the double copy construction can  be made manifest at the level of scattering amplitudes. More specifically, we proved a symmetric form of the Kawai-Lewellen-Tye relations which relate tree-level scattering amplitudes of bi-fluid quanta to products of tree-level colour ordered partial amplitudes of fluid quanta. Again in physical terminology, this is what one would mean by an ``on-shell'' formulation of the double copy.

The approach that we have followed has the virtue of giving us the proper arena for understanding the algebraic origins and different incarnations of the double copy procedure. It is crucial to notice that, although there are a number of features peculiar to the non-abelian Navier-Stokes equation, the constructions that we have presented are completely general and not limited to this example. This is particularly apparent in our proof of the Kawai-Lewellen-Tye relations, which does not rely on any other properties of the double copy Berends-Giele currents than those provided by the ``factorisation'' with respect to the colour-stripped Berends-Giele map. We therefore expect that results analogous to those described here in connection to the non-abelian and tensor Navier-Stokes equations can be obtained for other theories as well. Natural candidates include the self-dual sectors of Yang-Mills and gravity in the light-cone formulation \cite{Campiglia:2021srh} and  topologically massive $3$-dimensional Yang-Mills theory \cite{Moynihan:2021rwh}. Work along these lines is in progress.



\providecommand{\href}[2]{#2}\begingroup\raggedright\endgroup

\end{document}